\documentclass[journal,twocolumn]{IEEEtran}
\usepackage{amsfonts}%
\usepackage{amssymb}%
\usepackage{graphicx}
\usepackage{bbm}
\usepackage{jeffe}
\usepackage{epsfig}
\usepackage{subfigure}
\usepackage{amsmath,color}
\usepackage[all]{xy}
\usepackage{float}
\usepackage{color}

\newtheorem{them}{Theorem}
\newtheorem{defn}{Definition}

\newtheorem{lem}{Lemma}
\newtheorem{pro}{Proposition}
\newtheorem{remark}{Remark}

\def\d{ {\rm d  }}

\def\T{^{\rm\tiny T}}

\title{\Large \textbf{Cooperative Set Aggregation
for Multiple Lagrangian Systems}}

\author{Ziyang Meng, Tao Yang, Guodong Shi, Dimos V. Dimarogonas, \\ Yiguang Hong, and Karl H. Johansson
\thanks{Z. Meng, T. Yang, G. Shi, D. V. Dimarogonas, and K. H. Johansson are with ACCESS Linnaeus Centre, School of Electrical Engineering,
Royal Institute of Technology, Stockholm 10044, Sweden. Y. Hong is with Key Laboratory of Systems and Control, Institute of Systems Science, Chinese Academy of Sciences, Beijing 100190, China. Email: {\tt\small$\{$ziyangm, taoyang, guodongs, dimos, kallej$\}$@kth.se, yghong@iss.ac.cn.} Corresponding author: Z. Meng. Tel. +46-722-839377.}
\thanks{This work has been supported in part
by the Knut and Alice Wallenberg Foundation and the Swedish Research
Council.}}

\begin{document}
\maketitle

\begin{abstract}
In this paper, we study the cooperative set tracking problem for a group of Lagrangian systems. Each system observes a convex set as its local target. The intersection of these local sets is the group aggregation target. We first propose a control law based on each system's own target sensing and information exchange with neighbors. With necessary connectivity for both cases of fixed and switching communication graphs, multiple Lagrangian systems are shown to achieve rendezvous on the intersection of all the local target sets while the vectors of generalized coordinate derivatives are driven to zero. Then, we introduce the collision avoidance control term into set aggregation control to ensure group dispersion. By defining an ultimate bound on the final generalized coordinate between each system and the intersection of all the local target sets, we show that multiple Lagrangian systems approach a bounded region near the intersection of all the local target sets while the collision avoidance is guaranteed during the movement. In addition, the vectors of generalized coordinate derivatives of all the mechanical systems are shown to be driven to zero. Simulation results are given to validate the theoretical results.
\end{abstract}

\section{Introduction}

Along with the rapid development of coordination of multi-agent systems (see {\em e.g.,} \cite{VicsekEtAl95,LinBrouckeFrancis04,Zavlanos_IRO08,Hollinger_IJRR2010,Bullo_IEEE2011,Cortes_IRO2012,Rodriguez_RAS13,Oriolo_IJRR2013}), the study on the distributed control of multiple Lagrangian systems has
attracted extensive attention during the last decade. Compared with the single integrator dynamics, a Lagrangian model can be used to describe mechanical
systems, such as mobile robots, autonomous vehicles, robotic
manipulators, and rigid bodies. Therefore, the study on the distributed control of multiple Lagrangian systems is more applicable to the applications including spacecraft formation flying and relative attitude keeping and control of multiple unmanned aerial vehicles, just named a few.

The key idea of distributed control is to realize a collective task for the overall system by using only neighboring information exchange \cite{JadbabaieLinMorse03,SaberFaxMurray07_IEEE,CaoMing_TAC08,Cortes_IRO2009,Do_RAS13,Bullo_IRO12}. Such an algorithm relies on a setting that communication units are equipped for each individual system and thus a natural issue is communication link failure. Therefore, the analysis on the validness of distributed algorithm over a switching communication graph was investigated. Both continuous-time and discrete-time models were studied and many deep understanding was obtained for linear models \cite{JadbabaieLinMorse03,SaberMurray04,RenBeard05_TAC}. Nonlinear multi-agent dynamics has also drawn much attention \cite{Moreau_TAC05,LinZhiyun_SIAM07} since in many practical problems the node dynamics is naturally nonlinear, e.g., Vicsek's model and the Kuramoto's model \cite{VicsekEtAl95,Bullo_PNAS13}.

For the coordination problem of multiple Lagrangian systems, the author of \cite{Ren09_IJC} proposed distributed model-independent consensus algorithms for multiple Lagrangian systems in a leaderless setting. The case of time-varying leader was studied in
\cite{Chung_IRO09}, where the nonlinear contraction analysis was introduced to obtain
globally exponential convergence results. The connectedness maintenance problem was studied for multiple nonholonomic robotics
in \cite{Dimos_IRO2008}
and finite-time cooperative tracking algorithms were presented in \cite{Khoo_ITM09}
over graphs that are quasi-strongly connected.
Distributed containment control was proposed in \cite{MeiJie_Automatica12} and a sliding mode based strategy was introduced to estimate the leaders' generalized coordinate derivative information.
A similar problem was also studied in \cite{ChenGang_SMC11}, where continuous control algorithms were proposed to guarantee cooperative tracking with bounded errors. The authors of \cite{MengLinRen_SCL12} established containment, group dispersion and group cohesion behaviors for multiple Lagrangian systems, where both the cases of constant and time-varying leaders' velocities were considered. In addition, the applications of the coordination algorithms on shape control and  robotic manipulator synchronization were given, respectively, in \cite{Cheah_Automatica12} and \cite{Nikhil_IRO12}.

In this work, we focus on the cooperative set tracking problem of multiple Lagrangian systems. The set target is used to describe a common region for all the systems and each system has access to only the constrained information on this common set target. We first propose a control guaranteeing the set aggregation for all the systems over fixed graphs. Then we extend the result to the case of switching graphs and that of the collision avoidance requirements. The main contributions of our results are as follows:
\begin{itemize}
\item  A cooperative set tracking control is proposed for multiple Lagrangian systems. It is shown that under a general connectivity assumption for both fixed and switching graphs, multiple Lagrangian systems achieve rendezvous on the intersection of all the local target sets while the vectors of generalized coordinate derivatives are driven to zero.

\item Collision avoidance is guaranteed during the movement. In addition, we show that multiple Lagrangian systems approach a bounded region near the intersection of all the local target sets while the vectors of generalized coordinate derivatives are driven to zero.
\end{itemize}

The remainder of the paper is organized as follows. In
Section \ref{sec:prim}, we
give some basic notations and definitions on convex analysis, graph theory, Dini derivatives, and state the problem definition. A result for fixed interaction graphs is given in Section \ref{sec:fixed-1}. Then the cases with switching graphs and collision avoidance requirements are discussed in Sections \ref{sec:switching1} and \ref{sec:collision}. A brief concluding remark is given in
Section \ref{sec:conclusion}.

\section{Preliminaries}\label{sec:prim}

In this section, we introduce some mathematical preliminaries on convex analysis \cite{Aubin1991}, graph theory \cite{Godsil_Book}, and Dini derivatives \cite{Filippov_Book}. We also state the problem definition of this paper.

\subsection{Convex analysis} \label{sec:convex}

Denote $\|\cdot\|$ the Euclidean norm. For any nonempty set $\mathcal{S} \subseteq \mathbb{R}^m$, we use $\|x\|_{\mathcal{S}}=\inf_{y\in \mathcal{S}}\|x-y\|$ to describe the distance between $x\in \mathbb{R}^m$ and $\mathcal{S}$. A set $\mathcal{S}\subset \mathbb{R}^m$ is called convex if $(1-\zeta)x+\zeta y\in \mathcal{S}$ when $x\in\mathcal{S}$, $y\in\mathcal{S}$, and $0\leq \zeta\leq 1$.

Let $\mathcal{S}$ be a convex set. The convex projection of any $x\in \mathbb{R}^m$ onto $\mathcal{S}$ is denoted by $P_{\mathcal{S}}(x)\in\mathcal{S}$ satisfying $\|x-P_{\mathcal{S}}(x)\|=\|x\|_{\mathcal{S}}$. We also know that $\|x\|_{\mathcal{S}}^2$ is continuously differentiable for all $x\in \mathbb{R}^m$, and its gradient can be explicitly obtained by \cite{Aubin1991}:
\begin{equation}
\nabla\|x\|_{\mathcal{S}}^2=2(x-P_{\mathcal{S}}(x))\label{eq:set-deri}.
\end{equation}
Also, it is trivial to see that
\begin{equation}
(P_{\mathcal{S}}(x)-x)\T(P_{\mathcal{S}}(x)-y)\leq 0,~~ \forall y\in \mathcal{S}\label{eq:set-ineq}.
\end{equation}

\subsection{Graph theory}\label{sec:graph}

An undirected graph $\mathcal{G}$ consists of
a pair $(\mathcal{V}, \mathcal{E})$, where $
\mathcal{V}=\{1,2,\ldots,n\}$ is a finite, nonempty set of nodes and
$ \mathcal{E} \subseteq \mathcal{V}\times \mathcal{V}$ is a set of
unordered pairs of nodes.
An arc $\{j,i\}\in\mathcal{E}$ denotes that node $i,j$
can obtain each other' information mutually.
All neighbors of node $ i$ are denoted $\mathcal{N}_i :=
\{j:\{j,i\}\in \mathcal{E}\}$.
A path is a sequence of arcs of the form $\{i_1,i_2\},\{i_2,i_3\},\dots$. An undirected graph $\mathcal{G}$ is said to be connected if each node has an undirected path to any other node.

The adjacency matrix $A=[a_{ij}]\in\mathbb{R}^{n\times n}$ associated with the graph $\mathcal{G}$ is defined such that $a_{ij}$ is positive if $\{j,i\}\in\mathcal{E}$ and $a_{ij}=0$ otherwise. We also assume that $a_{ij}=a_{ji}$, for all $i,j\in\mathcal{V}$ for the undirected graph in this paper. The Laplacian matrix $L=[l_{ij}]\in\mathbb{R}^{n\times n}$ associated with $A$ is defined as $l_{ii}=\sum_{j\neq i}a_{ij}$ and $l_{ij}=-a_{ij}$, where $i\neq j$.

\subsection{Dini derivatives}

Let $D^+V(t,x(t))$ be the upper Dini derivative of $V(t,x(t))$ with respect to $t$, i.e.,
  \begin{equation*}
D^+V(t,x)=\lim_{\tau\rightarrow 0^+}\sup\frac{V(t+\tau,x(t+\tau))-V(t,x(t))}{\tau}.
  \end{equation*}
The following lemma is useful \cite{Danskin66}.
\begin{lem}\label{lem:Dini}
Suppose for each $i\in \mathcal{V}$, $V_i:\mathbb{R}\times \mathbb{R}^m\rightarrow \mathbb{R}$ is continuously differentiable. Let $V(t,x)=\max_{i\in \mathcal{V}} V_i(t,x)$, and let $\mathcal{V}_1(t)=\{i\in \mathcal{V}: V_i(t,x(t))=V(t,x(t))\}$ be the set of indices where the maximum is reached at time $t$. Then
\begin{equation*}
D^+V(t,x(t))=\max_{i\in \mathcal{V}_1(t)}\dot V_i(t,x(t)).
  \end{equation*}
\end{lem}

\subsection{Problem Definition}

Consider a network with $n$ agents labeled by $\mathcal{V}=\{1,2,\ldots,n\}$.
The dynamics of agent $i$ is described by the Lagrangian equations
\begin{equation}
M_i(q_i)\ddot q_i+C_i(q_i,\dot q_i)\dot q_i=\tau_i,~ i=1,2,\dots,n,
\label{eq:Lagrangian}
\end{equation}
where $q_i\in \mathbb{R}^{m}$ is the vector of generalized coordinates,
$M_i(q_i)\in \mathbb{R}^{m\times m}$ is the $m\times m$ inertia (symmetric) matrix,
$C_i(q_i,\dot q_i)\dot q_i$ is the Coriolis and centrifugal terms, and $\tau_i\in \mathbb{R}^{m}$ is the control force. The dynamics of a Lagrangian system satisfies the following properties \cite{Spong_Book2006}:

1. $M_i(q_i)$ is positive definite and is bounded for any $q_i\in \mathbb{R}^{m}$.

2. $\dot M_i(q_i)-2C_i(q_i,\dot q_i)$ is skew symmetric. 

3. $C_i(q_i,\dot q_i)$ is bounded with respect to $q_i$ and linearly bounded with respect to $\dot q_i$. More specifically, there is positive constant $k_{C}$ such that $\|C_i(q_i,\dot q_i)\|\leq k_{C}\|\dot q_i\|$.

\begin{figure}
 \begin{center}
\includegraphics[scale=0.4]{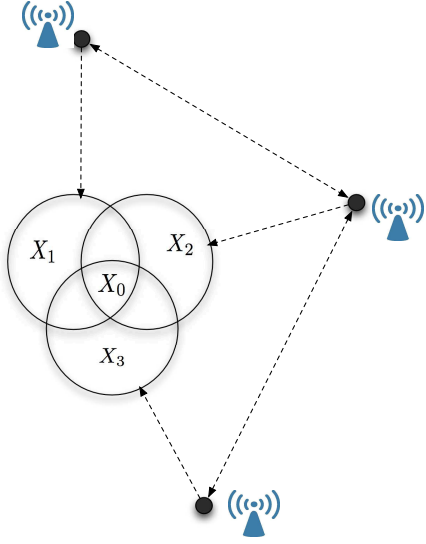}
 \end{center}
\caption{Cooperative set tracking problem}\label{fig:set-tracking}
\end{figure}

The objective of the group is to drive the multi-agent systems converge to a common region. Different from the existing works, we consider a set target objective instead of a point target objective.
The exact information of this common region is not available for all the agents. Instead, the constrained information for the region can be obtained by each agent using its' own sensor. We use set $\mathcal{X}_i$, $i\in \mathcal{V}$, to denote this available information for agent $i$. The global objective is to design a set aggregation control such that all the agents converge to the intersection of all $\mathcal{X}_i$, $i\in \mathcal{V}$, {\em i.e.,} $\mathcal{X}_0=\bigcap_i\mathcal{X}_i$. At each time, each agent observes the boundary points of its local available set and obtain the relative distance information between the local available set and itself. Also, the state information of each agent are exchanged by equipping each agent with simple and cheap communication unit. 	
The sketch map of the cooperative set tracking problem is presented in
Fig. \ref{fig:set-tracking}.

Based on these information, we first construct a set aggregation control for fixed graphs and switching graphs. Then, we consider set aggregation problem with collision avoidance control, where the relative distance information between different agents are used to derive a collision avoidance control.
In the following, we impose a standing assumption on set $\mathcal{X}_i$, $i\in \mathcal{V}$.

\vspace{2mm}
\noindent{\bf Assumption 1.} $\mathcal{X}_1, \mathcal{X}_2,\dots,\mathcal{X}_n$ are closed convex sets, and $\mathcal{X}_0=\bigcap_{i=1}^n \mathcal{X}_i$ is nonempty and bounded.
\vspace{2mm}

We introduce the following definition on cooperative set aggregation.
\begin{defn}\label{def1}
Multi-agent system \eqref{eq:Lagrangian} achieves cooperative set aggregation if
\begin{enumerate}
\item $\lim_{t\rightarrow \infty}\|q_i(t)\|_{\mathcal{X}_0}=0,\quad \forall i\in\mathcal{V}$,
\item $\lim_{t\rightarrow \infty}(q_i(t)-q_j(t))=0,\quad \forall i,j\in\mathcal{V}$,
\item $\lim_{t\rightarrow \infty}\dot q_i(t)=0,\quad \forall i\in\mathcal{V}$.
\end{enumerate}
\end{defn}
\begin{remark}
In fact, this set aggregation problem under convexity assumptions was a classical problem in optimization, where projected consensus algorithm was a standard solution \cite{Gubin67}. This algorithm was then generalized to distributed versions via consensus dynamics in \cite{Nedic_TAC10,Guodong_TAC13}. However, all these existing algorithms are designed for agents with first-order dynamics, and are therefore not applicable to the problem studied in this paper.
\end{remark}
In order to solve this problem, we construct the following algorithm
\begin{equation}
\tau_i=\tau^{non}_i+\tau^v_{i}+\tau^s_{i}+\tau^{in}_{i}+\tau^{a}_{i},\label{eq:control-fund}
\end{equation}
where $\tau^{non}_i$ represents nonlinear damping, $\tau^v_{i}$ represents generalized coordinate derivative damping, $\tau^s_{i}$ represents self-target tracking, $\tau^{in}_{i}$ represents inter coordination of different systems, and $\tau^{a}_{i}$ represents the collision avoidance control. We next specify the designs of different control terms in different cases.

\section{Cooperative aggregation over fixed communication graphs}\label{sec:fixed-1}

Let an undirected graph $\mathcal{G}=(\mathcal{V},\mathcal{E})$ define the communication graph.
Moreover, recall that $j$ is a neighbor of $i$ when $\{j,i\}\in \mathcal{E}$, and $\mathcal{N}_i$ represents the set of agent $i$'s neighbors.
The following control law is proposed for all $i\in\mathcal{V}$:
\begin{equation}
\tau_i=\underbrace{-k\dot q_i}_{\tau^v_{i}}\underbrace{-(q_i-P_{\mathcal{X}_i}(q_i))}_{\tau^s_{i}}\underbrace{-\sum_{j\in \mathcal{N}_i}a_{ij}(q_i-q_j)}_{\tau^{in}_{i}},\label{eq:control1}
\end{equation}
where $k>0$ denotes generalized coordinate derivative damping, $a_{ij}>0$ for all $i,j$ marks the strength of the information flow between $i$ and $j$.

\begin{them}\label{thm1}
Suppose that Assumption 1 hold. The multi-agent system \eqref{eq:Lagrangian} with \eqref{eq:control1} achieves set aggregation in the sense of Definition \ref{def1} if the fixed communication graph $\mathcal{G}$ is connected.
\end{them}

\proof
Consider the following Lyapunov function:
\begin{align*}
V=&~\frac{1}{2}\sum_{i=1}^n\dot q_i\T M_i(q_i) \dot q_i+\frac{1}{4}\sum_{i=1}^n\sum_{j\in \mathcal{N}_i}a_{ij}\|q_i-q_j\|^2
\\&+\frac{1}{2}\sum_{i=1}^n\|q_i-P_{\mathcal{X}_i}(q_i)\|^2.
\end{align*}
The derivative of $V$ along \eqref{eq:Lagrangian} with \eqref{eq:control1} is
\begin{align*}
\dot V=&~\sum_{i=1}^n\dot q_i\T \left(\frac{1}{2}\dot M_i(q_i) \dot q_i+ M_i(q_i)\ddot q_i\right)
\\&\!+\!\frac{1}{2}\sum_{i=1}^n\sum_{j\in \mathcal{N}_i}a_{ij}
 (q_i-q_j)\T
(\dot q_i-\dot q_j)\!+\!\!\sum_{i=1}^n\dot q_i\T(q_i-P_{\mathcal{X}_i}(q_i))
\\=&~\sum_{i=1}^n\dot q_i\T \left(-k\dot q_i-\sum_{j\in \mathcal{N}_i}a_{ij}(q_i-q_j)-(q_i-P_{\mathcal{X}_i}(q_i))\right)
\\&+\sum_{i=1}^n\dot q_i\T\sum_{j\in \mathcal{N}_i}a_{ij}(q_i-q_j)
+\sum_{i=1}^n\dot q_i\T(q_i-P_{\mathcal{X}_i}(q_i))
\\=&~-k\sum_{i=1}^n\|\dot q_i\|^2
\\ \leq &~0,
\end{align*}
where we have used \eqref{eq:set-deri} to derive the first equality and the fact that $a_{ij}=a_{ji}$ and the second property of Lagrangian dynamics to derive the second equality.

Therefore, based on LaSalle's Invariance Principle (Theorem 4.4 of \cite{Khalil_book}), we know that every solution of \eqref{eq:Lagrangian} with \eqref{eq:control1} converges to largest invariant set in $\mathcal{M}$, where $\mathcal{M}=\{q_i\in \mathbb{R}^m,\dot q_i\in \mathbb{R}^m,~\forall i\in \mathcal{V} ~|~ \dot q_i=0,\forall i\in \mathcal{V}\}$. Let $q_i(t)$, $\dot q_i(t)$, $\forall i\in \mathcal{V}$ be a solution that belongs to $\mathcal{M}$:
\begin{align*}
\dot q_i\equiv 0,~~\forall i\in \mathcal{V} \Rightarrow \ddot q_i\equiv 0,~~\forall i\in \mathcal{V}.
\end{align*}
It then follows from \eqref{eq:Lagrangian} and \eqref{eq:control1} that for all $ i\in \mathcal{V}$,
\begin{align*}
\tau_i\equiv 0 \Rightarrow \sum_{j\in \mathcal{N}_i}a_{ij}(q_i-q_j)+(q_i-P_{\mathcal{X}_i}(q_i))\equiv 0.
\end{align*}
Pick any $q_0\in \mathcal{X}_0$. Such a $q_0$ exists due to Assumption 1. Thus, it follows that for all $i\in \mathcal{V}$,
\begin{align*}
(q_i-q_0)\T\sum_{j\in \mathcal{N}_i}a_{ij}(q_i-q_j)+(q_i-q_0)\T&(q_i-P_{\mathcal{X}_i}(q_i))\equiv 0.
\end{align*}
We then know that
$$
\sum_{i=1}^n(q_i-q_0)\T\sum_{j\in \mathcal{N}_i}a_{ij}(q_i-q_j)
+\sum_{i=1}^n(q_i-q_0)\T(q_i-P_{\mathcal{X}_i}(q_i))\equiv 0.
$$
It follows that $\sum_{i=1}^n(q_i-q_0)\T\sum_{j\in \mathcal{N}_i}a_{ij}(q_i-q_j)=\frac{1}{2}\sum_{i=1}^n\sum_{j\in \mathcal{N}_i}a_{ij}\|q_i-q_j\|^2\geq 0$ by noting that $a_{ij}=a_{ji}$. Also, we know from \eqref{eq:set-ineq} that  $(P_{\mathcal{X}_i}(q_i)-q_0)\T(q_i-P_{\mathcal{X}_i}(q_i))\geq 0$. It then follows that
\begin{align*}
(q_i-q_0)\T(q_i-P_{\mathcal{X}_i}(q_i))=&~\|q_i-P_{\mathcal{X}_i}(q_i)\|^2
\\&+(P_{\mathcal{X}_i}(q_i)-q_0)\T(q_i-P_{\mathcal{X}_i}(q_i))
\\ \geq &~\|q_i-P_{\mathcal{X}_i}(q_i)\|^2
\\ \geq &~0.
\end{align*}
This shows that $\sum_{i=1}^n(q_i-q_0)\T\sum_{j\in \mathcal{N}_i}a_{ij}(q_i-q_j)\equiv 0$, and $\|q_i-P_{\mathcal{X}_i}(q_i)\|\equiv 0$, $\forall i\in \mathcal{V}$.
Therefore, we know that $\lim_{t\rightarrow\infty}(q_i(t)-P_{\mathcal{X}_i}(q_i(t)))=0 $, $\forall i\in \mathcal{V}$, and $\lim_{t\rightarrow\infty}(q_i(t)-q_j(t))=0$, $\forall i,j\in \mathcal{V}$. This shows that set aggregation is achieved in the sense of Definition \ref{def1}.
\endproof

In the following, we investigate the problem under switching communication graphs and collision avoidance, respectively, where the analysis becomes much more challenging.

\section{Cooperative aggregation over switching communication graphs}\label{sec:switching1}

One issue of introducing the communication unit is the possible communication link failure. The link failure becomes even more important when we consider the real applications including controlling multiple autonomous vehicles in the environments with limited power. Therefore, it is necessary to consider the case of switching communication graph. We associate the switching communication topology with a time-varying graph $\mathcal{G}_{\sigma(t)}=(\mathcal{V},\mathcal{E}_{\sigma(t)})$, where $\sigma:[t_0,+\infty)\rightarrow \mathcal{P}$ is a piecewise constant function and $\mathcal{P}$ is finite set of all possible graphs. $\mathcal{G}_{\sigma(t)}$ remains constant for $t\in [t_l,t_{l+1})$, $l=0,1,\dots$ and switches at $t=t_l$, $l=1,\dots$. In addition, we assume that $\inf_{l}(t_{l+1}-t_l)\geq \tau_d>0$, $l=1,\dots$, where $\tau_d$ is a constant and this dwell time assumption is extensively used in the analysis of switched systems \cite{Liberzon_ICSM1999}. The joint graph of $\mathcal{G}_{\sigma(t)}$ during time interval $[t_1,t_2)$ is defined by $\mathcal{G}_{\sigma(t)}([t_1,t_2))=\bigcup_{t\in [t_1,t_2)}\mathcal{G}(t)=(\mathcal{V},\bigcup_{t\in [t_1,t_2)}\mathcal{E}(t))$. Moreover, $j$ is a neighbor of $i$ at time $t$ when $\{j,i\}\in \mathcal{E}_{\sigma(t)}$, and $\mathcal{N}(\sigma(t))$ represents the set of agent $i$'s neighbors at time $t$.
\begin{defn}
$\mathcal{G}_{\sigma(t)}$ is uniformly jointly connected if there exists a constant $T>0$ such that $\mathcal{G}([t,t+T))$ is connected for any $t\geq t_0$.
\end{defn}

The existence of the switching communication graph complexes the problem significantly. In order to simplify the problem, we assume that the exact information of system dynamical parameters are available and propose the following control
\begin{align}
\tau_i=&~\underbrace{C_i(q_i,\dot q_i)\dot q_i}_{\tau_i^{non}}\underbrace{-kM_i(q_i)\dot q_i}_{\tau_i^{v}}\underbrace{-M_i(q_i)(q_i-P_{\mathcal{X}_i}(q_i))}_{\tau_i^{s}}
\notag\\&~\underbrace{-M_i(q_i)\sum_{j\in \mathcal{N}_i(\sigma(t))}a_{ij}(t)(q_i-q_j)}_{\tau_i^{in}},\qquad \forall i\in\mathcal{V},\label{eq:control2}
\end{align}
where $k>0$ denotes generalized coordinate derivative damping, and continuous function $a_{ij}(t)>0$ is the weight of arc $\{j,i\}$ for $i,j\in \mathcal{V}$ at $t$. We also assume that $a_{ij}(t)$ satisfies the following condition:

\vspace{2mm}
\noindent{\bf Assumption 2.}
There exists constants $a^*>0$ and $a_*>0$ such that for all $i,j\in \mathcal{V}$,
\begin{align*}
a_*\leq a_{ij}(t)\leq a^*,\quad t\in \mathbb{R}^+.
\end{align*}
\vspace{2mm}

Closed-loop system of \eqref{eq:Lagrangian} and \eqref{eq:control2} is
\begin{equation}
\ddot q_i=-k\dot q_i-\sum_{j\in \mathcal{N}_i(\sigma(t))}a_{ij}(t)(q_i-q_j)-(q_i-P_{\mathcal{X}_i}(q_i)).\label{eq:closed1}
\end{equation}
We next focus on the closed-loop system \eqref{eq:closed1}.

\subsection{Local set aggregation}

We first present a proposition regarding the local set aggregation of the closed-loop system \eqref{eq:closed1}. Based on this proposition, we then show global set aggregation.

\begin{pro}\label{lem:invariant}
Suppose that Assumptions 1 and 2 hold and choose $k$ large enough. The states of the multi-agent system \eqref{eq:Lagrangian} with \eqref{eq:control2} achieve local target aggregation, {\em i.e.,} $\lim_{t\rightarrow \infty}\|q_i(t)\|_{\mathcal{X}_i}=0$, and $\lim_{t\rightarrow \infty}\dot q_i(t)=0$, for all $i\in\mathcal{V}$.
\end{pro}
\proof
By picking any $q_0\in \mathcal{X}_0$,
we propose the following Lyapunov function:
\begin{align*}
V=&~\frac{1}{2}\sum_{i=1}^n\dot q_i\T \dot q_i+\sum_{i=1}^n(q_i-q_0)\T \dot q_i+\frac{k}{2}\sum_{i=1}^n\|q_i-q_0\|^2
\\&+\frac{1}{2}\sum_{i=1}^n\|q_i-P_{\mathcal{X}_i}(q_i)\|^2,
\end{align*}
where we choose $k>1$ to guarantee $V$ is positive definite.
The derivative of $V$ along \eqref{eq:closed1} is
\begin{align*}
\dot V=&\sum_{i=1}^n\dot q_i\T \left(\!\!-k\dot q_i-\!\!\!\!\!\!\sum_{j\in \mathcal{N}_i(\sigma(t))}a_{ij}(t)(q_i-q_j)-(q_i-P_{\mathcal{X}_i}(q_i))\!\!\right)
\\&+\sum_{i=1}^n(q_i-q_0)\T
\left(-k\dot q_i-\sum_{j\in \mathcal{N}_i(\sigma(t))}a_{ij}(t)(q_i-q_j)\right.
\\&\left.-(q_i-P_{\mathcal{X}_i}(q_i))\!\!\frac{}{}\right)
+\sum_{i=1}^n\|\dot q_i\|^2+k \sum_{i=1}^n(q_i-q_0)\T \dot q_i
\\&+\sum_{i=1}^n\dot q_i\T(q_i-P_{\mathcal{X}_i}(q_i))
\\=&~-(k-1)\sum_{i=1}^n\|\dot q_i\|^2-\sum_{i=1}^n\dot q_i\T\sum_{j\in \mathcal{N}_i(\sigma(t))}a_{ij}(t)(q_i-q_j)
\\&-\sum_{i=1}^n(q_i-q_0)\T(q_i-P_{\mathcal{X}_i}(q_i))
-\sum_{i=1}^n(q_i-q_0)\T
\\&~\times\sum_{j\in \mathcal{N}_i(\sigma(t))}a_{ij}(t)(q_i-q_j)
\\ \leq &~-\left[
\begin{array}{cc} q(t) & \dot q(t) \end{array} \right]\left[
\begin{array}{cc} L_{\sigma(t)} & \frac{L_{\sigma(t)}}{2}
\\ \frac{L_{\sigma(t)}}{2} & (k-2)I_{n} \end{array} \right]\left[
\begin{array}{c} q(t) \\ \dot q(t) \end{array} \right]
\\&-\sum_{i=1}^n\|q_i(t)\|_{\mathcal{X}_i}^2-\sum_{i=1}^n\|\dot q_i(t)\|^2,
\end{align*}
where $q=[q_1\T,q_2\T,\dots,q_n\T]\T$, $q=[q_1\T,q_2\T,\dots,q_n\T]\T$, $L_{\sigma(t)}$ is Laplacian matrix associated with $\mathcal{G}_{\sigma(t)}$ at time $t$ defined in Section \ref{sec:graph}, and we have used the fact that $(q_i-q_0)\T(q_i-P_{\mathcal{X}_i}(q_i))\geq \|q_i-P_{\mathcal{X}_i}(q_i)\|^2$. It is trivial to show that $L_{p}$ is symmetric and
positive semi-definite, for all $p\in\mathcal{P}$.
Therefore, if $k$ is chosen such that $k>2+\frac{1}{4}\max_{p\in\mathcal{P}}\{\lambda_{\max}(L_{p})\}$, or equivalent, $k>2+\frac{(n-1)a^*}{2}$, we can show that $\left[
\begin{array}{cc} L_{p} & \frac{L_{p}}{2}
\\ \frac{L_{p}}{2} & (k-2)I_{n} \end{array} \right]$ is positive semi-definite, for all $p\in\mathcal{P}$. It then follows that
\begin{align}
\dot V\leq -\sum_{i=1}^n\|q_i\|_{\mathcal{X}_i}^2-\sum_{i=1}^n\|\dot q_i\|^2\leq 0.\label{eq:deri}
\end{align}
Therefore, $q_i$ and $\dot q_i$, $\forall i\in \mathcal{V}$ are bounded.
We also know that \eqref{eq:deri} implies that
\begin{align*}
\int_{t_0}^{\infty}\left(\sum_{i=1}^n\|q_i(t)\|_{\mathcal{X}_i}^2+\sum_{i=1}^n\|\dot q_i(t)\|^2\right)\d t\leq V(t_0)
\end{align*}
is bounded.
In addition, it follows that $
\frac{\d}{\d t}\left(\sum_{i=1}^n\|q_i(t)\|_{\mathcal{X}_i}^2+\sum_{i=1}^n\|\dot q_i(t)\|^2\right)
\\=~2\sum_{i=1}^n \left((q_i-P_{\mathcal{X}_i}(q_i))\T \dot q_i+\dot q_i\T\ddot q_i\right).$
Therefore, from \eqref{eq:closed1} and the facts that $q_i$ and $\dot q_i$, $\forall i\in \mathcal{V}$ are bounded, we know that $
\frac{\d}{\d t} \left(\sum_{i=1}^n
\|q_i(t)\|_{\mathcal{X}_i}^2+\sum_{i=1}^n\|\dot q_i(t)\|^2\right)$ is bounded $\forall t\in\mathbb{R}^+$. Then, based on Barbalat's lemma \cite{Khalil_book}, we can show that
$\sum_{i=1}^n\|q_i(t)\|_{\mathcal{X}_i}^2+\sum_{i=1}^n\|\dot q_i(t)\|^2\rightarrow 0$, as $t\rightarrow \infty$. Therefore, $\lim_{t\rightarrow \infty}\|q_i(t)\|_{\mathcal{X}_i}=0$, and $\lim_{t\rightarrow \infty}\dot q_i(t)=0$, for all $i\in\mathcal{V}$.
\endproof

\subsection{Global set aggregation}
Next, we rewrite the closed-loop system \eqref{eq:closed1} as
\begin{subequations}\label{eq:double1}
\begin{equation}
\dot q_i=\dot q_i,
\end{equation}
\begin{equation}
\ddot q_i=-k\dot q_i-\sum_{j\in \mathcal{N}_i(\sigma(t))}a_{ij}(t)(q_i-q_j)-(q_i-P_{\mathcal{X}_i}(q_i)).
\end{equation}
\end{subequations}
Define $x_i=q_i$, $x_{n+i}=q_i+\frac{2}{k}\dot q_i$, for all $i\in\mathcal{V}$. After some manipulations, \eqref{eq:double1} can be written as
\begin{subequations}
\begin{equation*}
\dot x_i=-\frac{k}{2}(x_i-x_{n+i}),~~i=1,2,\dots,n
\end{equation*}
\begin{align*}
\dot x_{n+i}=&-\frac{k}{2}(x_{n+i}-x_i)-\frac{2}{k}\sum_{j\in \mathcal{N}_i(\sigma(t))}a_{ij}(t)(x_{n+i}-x_{n+j})
\\&+\delta_i(t),~~i=1,2,\dots,n.
\end{align*}
\end{subequations}
where $\delta_i=\frac{4}{k^2}\sum_{j\in \mathcal{N}_i(\sigma(t))}a_{ij}(t)(\dot q_i-\dot q_j)-\frac{2}{k}(q_i-P_{\mathcal{X}_i}(q_i))$, for all $i\in\mathcal{V}$. Note that each entry of $x_i\in\mathbb{R}^m$ is decoupled, without loss of generality, we assume $m=1$ in the following analysis.

Define $\overline{\mathcal{V}}=\{1,2,\dots,2n\}$, and
$$\hbar(t)=\max_{i\in \overline{\mathcal{V}}}\{x_{i}(t)\},~~\ell(t)=\min_{i\in \overline{\mathcal{V}}}\{x_{i}(t)\}.$$
\begin{lem}\label{lem:max}
For all $t\geq t_0\geq 0$, it follows that
\begin{align*}
D^+\hbar(t)\leq \max_{i\in \mathcal{V}}\|\delta_i\|,\quad D^+\ell(t)\geq -\max_{i\in \mathcal{V}}\|\delta_i\|.
\end{align*}
\end{lem}

\proof

Let $\overline{\mathcal{V}}_1(t)$ be the set containing all the agents that reach the maximum at time $t$, i.e.,  $\overline{\mathcal{V}}_1(t)=\{i\in \overline{\mathcal{V}} | x_i(t)=\hbar(t)\}$. It then follows from Lemma \ref{lem:Dini} that
\begin{align*}
D^+\hbar(t)\leq& ~\max_{i\in \overline{\mathcal{V}}_1}\dot x_i(t)
\\ \leq&~\max\left\{~\max_{i\in \overline{\mathcal{V}}_1\bigcap\{1,2,\dots,n\}}\frac{k}{2}(x_{n+i}-x_i),\right.
\\& \max_{i\in \overline{\mathcal{V}}_1\bigcap\{n+1,n+2,\dots,2n\}}\frac{k}{2}(x_{i-n}-x_i)
\\&\left.\!\!+\frac{2}{k}\sum_{j\in \mathcal{N}_{i-n}(\sigma(t))}a_{(i-n)j}(t)(x_{n+j}-x_{i})+\delta_{i-n}(t) \!\! \right\}
\\ \leq&~ \max_{i\in\mathcal{V}(t)}\delta_i(t)
\\ \leq&~  \max_{i\in \mathcal{V}}\|\delta_i\|.
\end{align*}
Similarly, we have that $D^+\ell(t)\geq -\max_{i\in \mathcal{V}}\|\delta_i\|$.
\endproof

\begin{them}
Suppose that Assumptions 1 and 2 hold and choose $k$ large enough. The multi-agent system \eqref{eq:Lagrangian} with \eqref{eq:control2} achieves set aggregation in the sense of Definition \ref{def1} if the communication graph $\mathcal{G}_{\sigma(t)}$ is uniformly jointly connected.
\end{them}
\proof
It follows from Proposition \ref{lem:invariant} that $\lim_{t\rightarrow\infty}\delta_i(t)=0$, for all $i\in \mathcal{V}$.
This shows that for any $\varepsilon>0$, there exists $T_1(\varepsilon)>0$ such that
\begin{align*}
\|\delta_i(t)\|\leq \varepsilon,\quad \forall i\in \mathcal{V},\quad t\geq T_1.
\end{align*}
It also follows from Proposition \ref{lem:invariant} that $x_i(t)$, for all $i\in\overline{\mathcal{V}}$, is bounded for all $t\geq t_0\geq 0$. Therefore, $\hbar(T_1)$ and $\ell(T_1)$ are bounded.

The analysis of this theorem is motivated by \cite{Guodong_TAC13}. Define $\overline{T}=T+2\tau_d$.
We next focus on the time interval $t\in [T_1,T_1+N\overline{T}]$, where $N=n+1$. It follows from Lemma \ref{lem:max} that
\begin{align}\label{eq:max-less}
\hbar(t)\leq \hbar(T_1)+N\overline{T}\varepsilon,\quad \ell(t)\geq \ell(T_1)-N\overline{T}\varepsilon.
\end{align}
Without loss of generality, we assume that $\hbar(T_1)\neq \ell(T_1)$. Then, we consider a node $i_0\in \overline{\mathcal{V}}$, where we assume $i_0$ satisfies $x_{i_0}(T_1)\leq \frac{1}{2}\hbar(T_1)+\frac{1}{2}\ell(T_1)=\hbar(T_1)-\varepsilon_0$, and $\varepsilon_0=\frac{1}{2}(\hbar(T_1)-\ell(T_1))>0$. We first assume that $i_0\in\mathcal{V}$. It follows from \eqref{eq:max-less} that for all $t\in[T_1,T_1+N\overline{T}]$,
\begin{align*}
\dot x_{i_0}(t)\leq -\frac{k}{2}&\left(x_{i_0}(t)-\hbar(T_1)-N\overline{T}\varepsilon\right).
\end{align*}
By using Gronwall's inequality, we know that for all $t\in[T_1,T_1+N\overline{T}]$,
\begin{align}
 x_{i_0}(t)\leq &~x_{i_0}(T_1)e^{-\frac{k}{2}(t-T_1)}
 +(1- e^{-\frac{k}{2}(t-T_1)})
(\hbar(T_1)+N\overline{T}\varepsilon)
\notag \\ \leq &~ \hbar(T_1)- e^{-\frac{k}{2}(t-T_1)}\varepsilon_0+N\overline{T}\varepsilon
\notag \\ \leq &~ \hbar(T_1)- \overline{\varepsilon}_0+N\overline{T}\varepsilon, \label{eq:1}
\end{align}
where $\overline{\varepsilon}_0=e^{-kN\overline{T}/2}\varepsilon_0$.
It then follows that for all $t\in[T_1,T_1+N\overline{T}]$,
\begin{align*}
\dot x_{i_0+n}(t)\leq &~-\frac{k}{2}(x_{i_0+n}(t)-x_{i_0}(t))-\frac{2}{k}\left(\sum_{j=1}^{n}a_{i_0j}(t)\right)
\\&~\times(x_{i_0+n}(t)-\hbar(T_1)-N\overline{T}\varepsilon)+\varepsilon
\end{align*} It then follows from \eqref{eq:1} that
\begin{align*}
\dot x_{i_0+n}(t)\leq &~-\frac{k}{2}(x_{i_0+n}(t)-(\hbar(T_1)- \overline{\varepsilon}_0+N\overline{T}\varepsilon))
\\&~-\alpha(x_{i_0+n}(t)-\hbar(T_1)-N\overline{T}\varepsilon)+\varepsilon
\\ \!\leq &-\alpha_1(x_{i_0+n}(t)-(\hbar(T_1)- \frac{k}{2\alpha_1}\overline{\varepsilon}_0+N\overline{T}\varepsilon))+\varepsilon,
\end{align*}
where $\alpha=\frac{2}{k}(n-1)a^*$ and $\alpha_1=\frac{k}{2}+\frac{2}{k}(n-1)a^*$.
Thus, for all $t\in[T_1,T_1+N\overline{T}]$,
\begin{align*}
 x_{i_0+n}(t) \leq& ~ \hbar(T_1)-(1-e^{-\alpha_1(t-T_1)})\frac{k}{2\alpha_1}\overline{\varepsilon}_0
 +(N\overline{T}+\frac{2}{k})\varepsilon
\end{align*}
Thus, we know that
\begin{align*}
 x_{i_0+n}(T_1+\overline{T}) \leq &~ \hbar(T_1)- (1-e^{-\alpha_1\overline{T}})\frac{k}{2\alpha_1}\overline{\varepsilon}_0
  \\&+(N\overline{T}
+\frac{2}{k})\varepsilon
\end{align*}
Therefore, for all $t\in[T_1+\overline{T},T_1+N\overline{T}]$, it follows that
\begin{align*}
 x_{i_0+n}(t) \leq &~ \hbar(T_1)- (1-e^{-\alpha_1\overline{T}})\frac{k}{2\alpha_1}e^{-\alpha_1N\overline{T}} \overline{\varepsilon}_0
 \\&+(N\overline{T}+\frac{4}{k})\varepsilon
\end{align*}

Instead, if $i_0\in\{n+1,n+2,\dots,2n\}$,
it follows from \eqref{eq:max-less} that for all $t\in[T_1,T_1+N\overline{T}]$,
\begin{align*}
\dot x_{i_0}(t)\leq -\alpha_1\left(x_{i_0}(t)-\hbar(T_1)-N\overline{T}\varepsilon\right)
+\varepsilon.
\end{align*}
By using Gronwall's inequality, we know that for all $t\in[T_1,T_1+N\overline{T}]$,
\begin{align*}
 x_{i_0}(t)\leq~ \hbar(T_1)- e^{-\alpha_1N\overline{T}}\varepsilon_0+(N\overline{T} +\frac{2}{k})\varepsilon,
\end{align*}
Therefore, we know that there exists a $i_0\in \mathcal{V}$ satisfying
\begin{align*}
 x_{i_0+n}(t)\leq~ \hbar(T_1)- \overline{\varepsilon}_1+(N\overline{T}+\frac{4}{k})\varepsilon
\end{align*}
for all $t\in[T_1+\overline{T},T_1+N\overline{T}]$,
where $\overline{\varepsilon}_1=(1-e^{-\alpha_1\overline{T}})\frac{k}{2\alpha_1}e^{-\alpha_1N\overline{T}} e^{-\frac{k}{2}N\overline{T}} \varepsilon_0$.

Next, based on the fact that $\mathcal{G}_{\sigma(t)}$ is uniformly jointly connected, we know that $i_0+n$ is jointly connected to other node $j\in \{n+1,n+2,\dots,2n\}/(i_0+n)$ during the time interval $[T_1+\overline{T},T_1+\overline{T}+T]$. Therefore, there exists $[T_2,T_2+\tau_d]\subset[T_1+\overline{T},T_1+2\overline{T}]$ such that there is a $j\in \{n+1,n+2,\dots,2n\}/(i_0+n)$ being the neighbor of $i_0+n$ during $[T_2,T_2+\tau_d]$. Denote this node as $i_1+n$. We next establish an upper bound for $x_{i_1+n}(t)$. For any $t\in[T_2,T_2+\tau_d]$, we consider two cases:

Case I: $x_{i_1+n}(t)>x_{i_0+n}(t)$ for all $t\in[T_2,T_2+\tau_d]$. It then follows that
\begin{align*}
\dot x_{i_1+n}(t)\leq &~
-\left(\frac{k}{2}+\frac{2}{k}a^*(n-2)\right)\left(x_{i_1+n}
-\hbar(T_1)-N\overline{T}\varepsilon\right)
\\&~-\frac{2}{k}a_{i_1i_0}(t)\left(x_{i_1+n}(t)-x_{i_0+n}(t)\right)+\varepsilon
\\ \leq & ~ -\left(\frac{k}{2}+\frac{2}{k}a^*(n-2)\right)\left(x_{i_1+n}
-\hbar(T_1)-N\overline{T}\varepsilon\right)
\\&~-\frac{2}{k}a_*\left(x_{i_1+n}- (\hbar(T_1)-\overline{\varepsilon}_1+(N\overline{T}+\frac{4}{k})\varepsilon)\right)
\\&~+\varepsilon
\\ \leq & ~ -\alpha_2\left(x_{i_1+n}
-\hbar(T_1)+ \frac{2a_*}{k\alpha_2}\overline{\varepsilon}_1-(2N\overline{T}+\frac{4}{k})\varepsilon\right)
\\&~+\varepsilon,
\end{align*}
where $\alpha_2= \frac{k}{2}+\frac{2}{k}a^*(n-2)+\frac{2}{k}a_*$
It then follows that
\begin{align*}
x_{i_1+n}(T_2+\tau_d)\leq &~\hbar(T_1)-\left(1-e^{-\tau_d\alpha_2}\right)\frac{2a_*}{k\alpha_2}
\overline{\varepsilon}_1
\\&~+(2N\overline{T}+\frac{6}{k})\varepsilon
\end{align*}
Therefore, it follows that for all $t\in[T_2+\tau_d,T_1+N\overline{T}]$,
\begin{align*}
x_{i_1+n}(t)\leq &~\hbar(T_1)
-e^{-N\overline{T}\alpha_1}\left(1-e^{-\tau_d\alpha_2}\right)\frac{2a_*}{k\alpha_2}
\overline{\varepsilon}_1
\\&~+(2N\overline{T}+\frac{8}{k})\varepsilon.
\end{align*}

Case II: there exists a $t^*\in[T_2,T_2+\tau_d]$ such that $x_{i_1+n}(t^*)\leq x_{i_0+n}(t^*)$. It thus follows that
\begin{align*}
x_{i_1+n}(t^*)\leq \hbar(T_1)- \overline{\varepsilon}_1+(N\overline{T}+\frac{4}{k})\varepsilon.
\end{align*}
Therefore, we know that for all $t\in [t^*,T_1+N\overline{T}]$,
\begin{align*}
\dot x_{i_1+n}(t)\leq &~
-\alpha_1\left(x_{i_1+n}(t)
-\hbar(T_1)-N\overline{T}\varepsilon\right)
+\varepsilon.
\end{align*}
It then follows that
\begin{align*}
x_{i_1+n}(t)\leq &~\hbar(T_1)-e^{-N\overline{T}\alpha_1}\overline{\varepsilon}_1
+(N\overline{T}+\frac{6}{k})\varepsilon.
\end{align*}

Combing Cases I and II, we know that for all $t\in[T_2+\tau_d,T_1+N\overline{T}]$,
\begin{align*}
x_{i_1+n}(t)\leq &~\hbar(T_1)-\overline{\varepsilon}_2+(2N\overline{T}+\frac{8}{k})\varepsilon,
\end{align*}
where $\overline{\varepsilon}_2=e^{-N\overline{T}\alpha_1}
\left(1-e^{-\tau_d\alpha_2}\right)\frac{2a_*}{k\alpha_2}
\overline{\varepsilon}_1$.

By repeating the above process during the time interval $[T_1+\gamma \overline{T},T_1+(\gamma+1)\overline{T}]$, $\gamma=2,3,\dots n-1$, we can show that for all $i\in \mathcal{V}$ and $t\in [T_1+n\overline{T},T_1+N\overline{T}]$,
\begin{align*}
x_{i+n}(t)\leq ~ \hbar(T_1)- \overline{\varepsilon}_{n} +(N\overline{T}+\frac{4}{k})n\varepsilon,
\end{align*}
where $\overline{\varepsilon}_{n}=e^{-N(n-1)\overline{T}\alpha_1}
\left(1-e^{-\tau_d\alpha_2}\right)^{n-1}\left(\frac{2a_*}{k\alpha_2}\right)^{n-1}
\overline{\varepsilon}_1$. Then, we know that for all $i\in \mathcal{V}$, and $t\in[T_1+n\overline{T},T_1+N\overline{T}]$,
\begin{align*}
\dot x_{i}(t)\leq &~-\frac{k}{2}(x_i(t)-x_{i+n}(t))
\\ \leq &~-\frac{k}{2}(x_i(t)-(\hbar(T_1)- \overline{\varepsilon}_{n} +(N\overline{T}+\frac{4}{k})n\varepsilon)).
\end{align*}
Therefore, it follows that for all $i\in \mathcal{V}$,
\begin{align*}
\hbar(T_1+N\overline{T})\leq ~\hbar(T_1)-\overline{\varepsilon}_{n+1}+(N\overline{T}+\frac{4}{k})n\varepsilon,
\end{align*}
where $\overline{\varepsilon}_{n+1}=(1-e^{-\alpha_1N\overline{T}})\overline{\varepsilon}_{n}$.

Otherwise, if $i_0$ satisfies $x_{i_0}(T_1)\geq \frac{1}{2}\hbar(T_1)+\frac{1}{2}\ell(T_1)=\ell(T_1)+\varepsilon_0$, we can similarly show that
\begin{align*}
\ell(T_1+N\overline{T})\geq ~\ell(T_1)+\overline{\varepsilon}_{n+1}+(N\overline{T}+\frac{4}{k})n\varepsilon.
\end{align*}

It then follows that
\begin{align*}
\hbar(T_1+N\overline{T})\!-\!\ell(T_1+N\overline{T})\!\leq& ~ \hbar(T_1)\!-\!\ell(T_1)-\overline{\varepsilon}_{n+1}
\\&~+(N\overline{T}+\frac{4}{k})n\varepsilon.
\end{align*}

This implies that
\begin{align*}
\hbar(T_1+N\overline{T})\!-\!\ell(T_1+N\overline{T})\!\leq&~(1-\delta) (\hbar(T_1)\!-\!\ell(T_1))
\\&~+(N\overline{T}+\frac{4}{k})n\varepsilon,
\end{align*}
where $\delta=e^{-N(n-1)\overline{T}\alpha_1}\left(1-e^{-\tau_d\alpha_2}\right)^{n-1}\left(\frac{2a_*}{k\alpha_2}\right)^{n-1}
(1-e^{-\alpha_1\overline{T}})\frac{k}{4\alpha_1}e^{-(\alpha_1+\frac{k}{2})N\overline{T}}$. Then, for all $\xi=1,2,\dots,$, we obtain that
\begin{align*}
\hbar(T_1+\xi N\overline{T})-\ell(T_1+\xi N\overline{T})
\leq& ~(1-\delta)^\xi (\hbar(T_1)-\ell(T_1))\\&~+\sum_{j=0}^{\xi-1}(1-\delta)^{j}(N\overline{T}+\frac{4}{k})n\varepsilon
\\ \leq&~(1-\delta)^\xi (\hbar(T_1)-\ell(T_1))\\&~+\frac{(N\overline{T}+\frac{4}{k})n\varepsilon}{\delta}.
\end{align*}
Thus,
\begin{align*}
\lim_{\xi\rightarrow\infty}\sup(\hbar(T_1+\xi N\overline{T})&-\ell(T_1+\xi N\overline{T})) \leq \frac{(N\overline{T}+\frac{4}{k})n\varepsilon}{\delta},
\end{align*}
which shows that $\lim_{\xi\rightarrow\infty}\sup(\hbar(T_1+\xi N\overline{T})-\ell(T_1+\xi N\overline{T})) =0$
by choosing $\varepsilon$ sufficiently small. Therefore, we know that $\lim_{t\rightarrow\infty}(x_i(t)-x_j(t))=0$, for all $i,j\in \overline{\mathcal{V}}$. This implies that $\lim_{t\rightarrow\infty}(q_i(t)-q_j(t))=0$, for all $i,j\in \mathcal{V}$. This shows that set aggregation is achieved in sense of Definition \ref{def1}.
\endproof

\subsection{Simulation Verifications}

We now  use numerical simulations to validate the effectiveness of the theoretical results obtained in Sections \ref{sec:fixed-1} and \ref{sec:switching1}. We assume that there are eight agents ($n=8$) in the group. The system dynamics are given by \cite{Spong_Book2006},
\begin{align*}
\left[
\begin{array}{cc}
M_{11,i}& M_{12,i}\\
M_{21,i}& M_{22,i}\\
\end{array}
\right]\left[
\begin{array}{c}
\ddot q_{ix}\\
\ddot q_{iy} \\
\end{array}
\right]+&\left[
\begin{array}{cc}
C_{11,i}& C_{12,i}\\
C_{21,i}& C_{22,i} \\
\end{array}
\right]\left[
\begin{array}{c}
\dot q_{ix}\\
\dot q_{iy} \\
\end{array}
\right]
\\&=\left[
\begin{array}{c}
\tau_{ix}\\
\tau_{iy} \\
\end{array}
\right],i=1,2,\dots,8,
\end{align*}
where $M_{11,i}=\theta_{1i}+2\theta_{2i}\cos q_{iy}$, $M_{12,i}=M_{21,i}=\theta_{3i}+\theta_{2i}\cos q_{iy}$,
$M_{22,i}=\theta_{3i}$, $C_{11,i}=-\theta_{2i}\sin q_{iy}\dot q_{iy}$, $C_{12,i}=-\theta_{2i}\sin q_{iy}(\dot q_{ix}+\dot q_{iy})$, $C_{21,i}=\theta_{2i}\sin q_{iy}\dot q_{ix}$, $C_{22,i}=0$. We choose
$\theta_{1i}=1.301$, $\theta_{2i}=0.256$, $\theta_{3i}=0.096$, $i=1,2,\dots,8$.

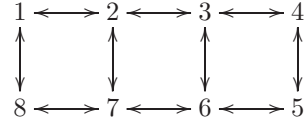
\begin{figure}[t]\centering
\begin{tabular} {c}
 \xymatrix{
1\ar@{<->}[r] \ar@{<->}[d]&  2\ar@{<->}[r] \ar@{<->}[d]& 3\ar@{<->}[d] & 4\ar@{<->}[l]\ar@{<->}[d]\\
8\ar@{<->}[r]           &  7\ar@{<->}[r]           & 6 \ar@{<->}[r]          & 5
} \end{tabular}\caption{The communication graph
$\mathcal{G}$ for Section \ref{sec:simulation1}} \label{fig:COMMTOP1}
\end{figure}

\subsubsection{Fixed communication graphs}\label{sec:simulation1}

we assume that the available local sets of all the agents are circles, where the radius of the circles are $3$ and circles given by $\mathcal{X}_1=(1.5+3\cos(\theta),1.5+3\sin(\theta))$, $\mathcal{X}_2=(-1.5+3\cos(\theta),-1.5+3\sin(\theta))$, $\mathcal{X}_3=(1.5+3\cos(\theta),1.5+3\sin(\theta))$, and $\mathcal{X}_4=(3\cos(\theta),-1.5+3\sin(\theta))$, $\mathcal{X}_5=(3\cos(\theta),-1.5+3\sin(\theta))$, $\mathcal{X}_6=(-1.5+3\cos(\theta),-1.5+3\sin(\theta))$, $\mathcal{X}_7=(1+3\cos(\theta),1+3\sin(\theta))$, and $\mathcal{X}_8=(-1.5+3\cos(\theta),-1.5+3\sin(\theta))$ with $\theta\in[0,2\pi)$.
The initial states of the agents are given by
$q_{1}(0) = [-8,8]\T$, $q_{2}(0) = [6.4,12]\T$, $q_{3}(0) = [-8,-8]\T$, $q_{4}(0) = [6,-8]\T$, $q_{5}(0) = [-8.8,-4]\T$, $q_{6}(0) = [4.8,-12]\T$, $q_{7}(0) = [-4,-8]\T$, $q_{8}(0) = [3.2,-12]\T$, $\dot q_{1}(0) = [-0.4,0.4]\T$, $\dot q_{2}(0) = [0.8,-0.8]\T$, $\dot q_{3}(0) = [2.8,-2.8]\T$, and $\dot q_{4}(0) = [1.6,-1.6]\T$, $\dot q_{5}(0) = [-1.2,0.8]\T$, $\dot q_{6}(0) = [1.6,-0.4]\T$, $\dot q_{7}(0) = [1.6,-2]\T$, and $\dot q_{8}(0) = [0.8,-0.8]\T$.
The control parameters
are chosen by $k=1$.
The communication graph
$\mathcal{G}$ is given in Fig. \ref{fig:COMMTOP1}.
Also, the weight of adjacency matrix $A$ of the generalized coordinates associated with
$\mathcal{G}$ is chosen to be $1$.

\begin{figure}
 \begin{center}
\includegraphics[scale=0.45]{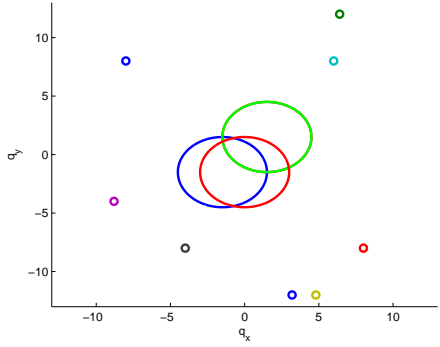}
\includegraphics[scale=0.45]{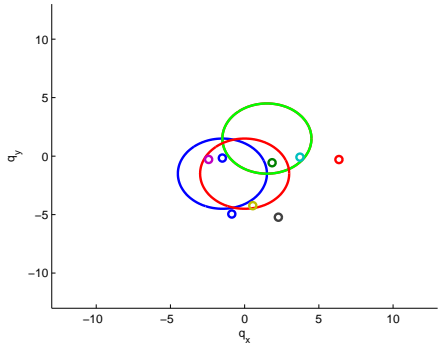}
\includegraphics[scale=0.45]{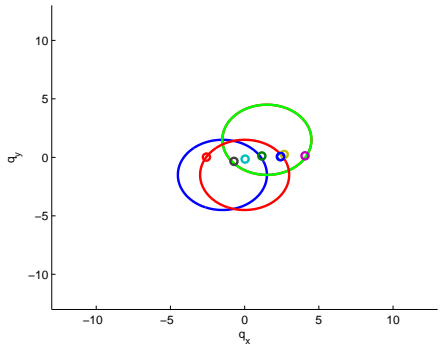}
\includegraphics[scale=0.45]{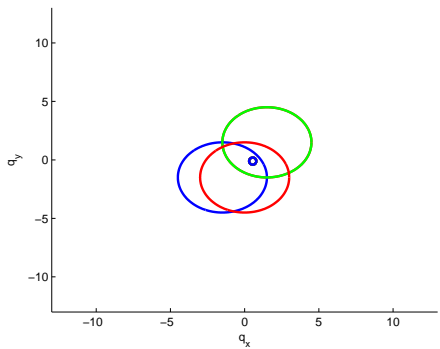}
 \end{center}
\caption{The trajectories of the generalized coordinates of the agents under control \eqref{eq:control1}. The circles denote the
generalized coordinates of the agents.}\label{fig:position1}
\end{figure}

\begin{figure}
 \begin{center}
\includegraphics[scale=0.5]{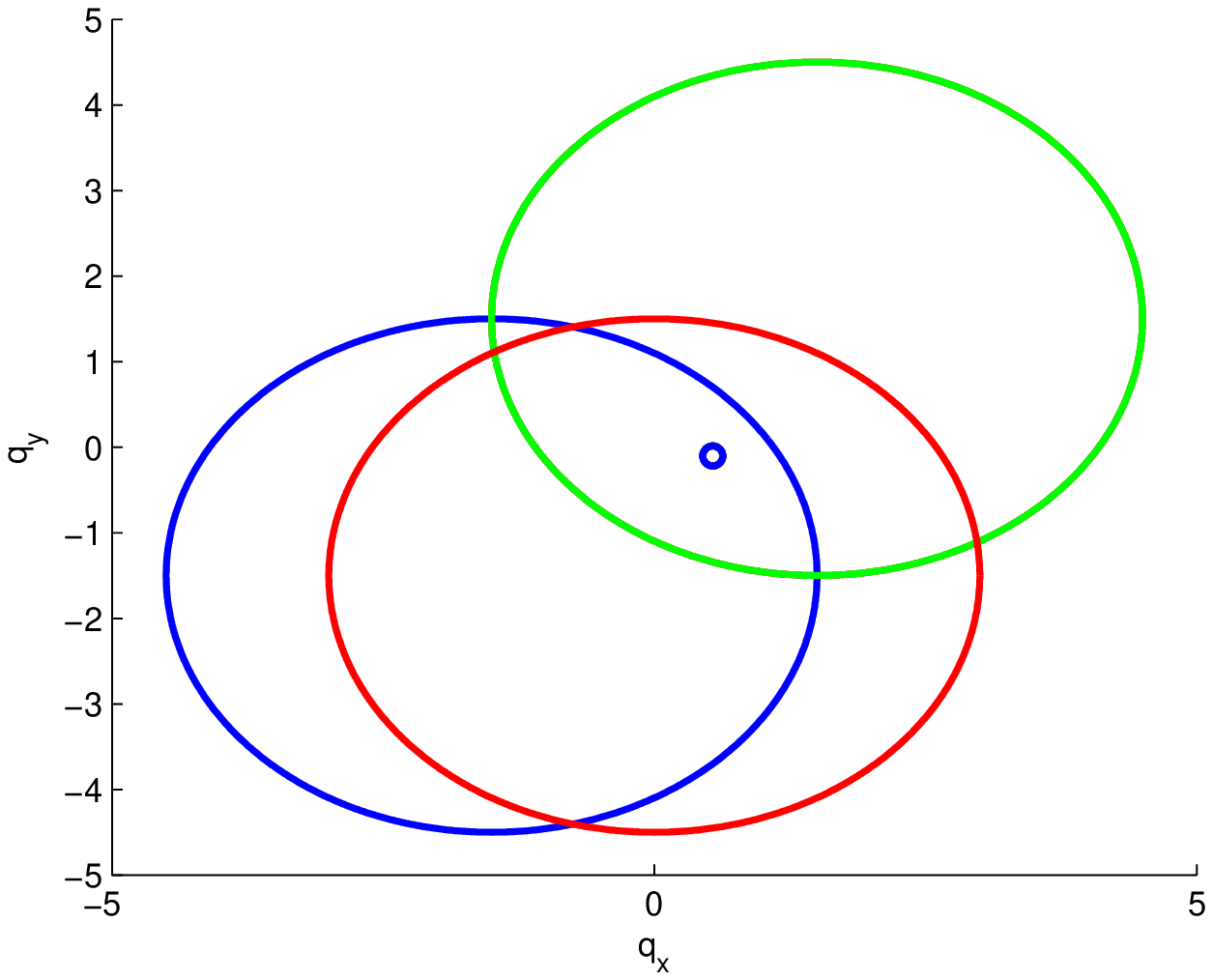}
 \end{center}
\caption{Zoom in of the final generalized coordinates of the agents under control \eqref{eq:control1}.}\label{fig:position1-zoomin}
\end{figure}

\begin{figure}
 \begin{center}
\includegraphics[scale=0.5]{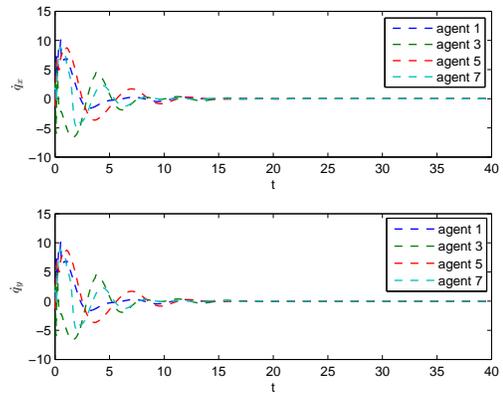}
 \end{center}
\caption{The trajectories of the generalized coordinate derivatives of the agents under control \eqref{eq:control1}.}\label{fig:velocity1}
\end{figure}

Under the control \eqref{eq:control1},
snapshots of generalized coordinates and trajectories of generalized coordinate derivatives of the agents
are shown in
Figs. \ref{fig:position1}, \ref{fig:position1-zoomin} and \ref{fig:velocity1}.
We see that set aggregation is achieved.

\begin{figure}
 \begin{center}
\includegraphics[scale=0.45]{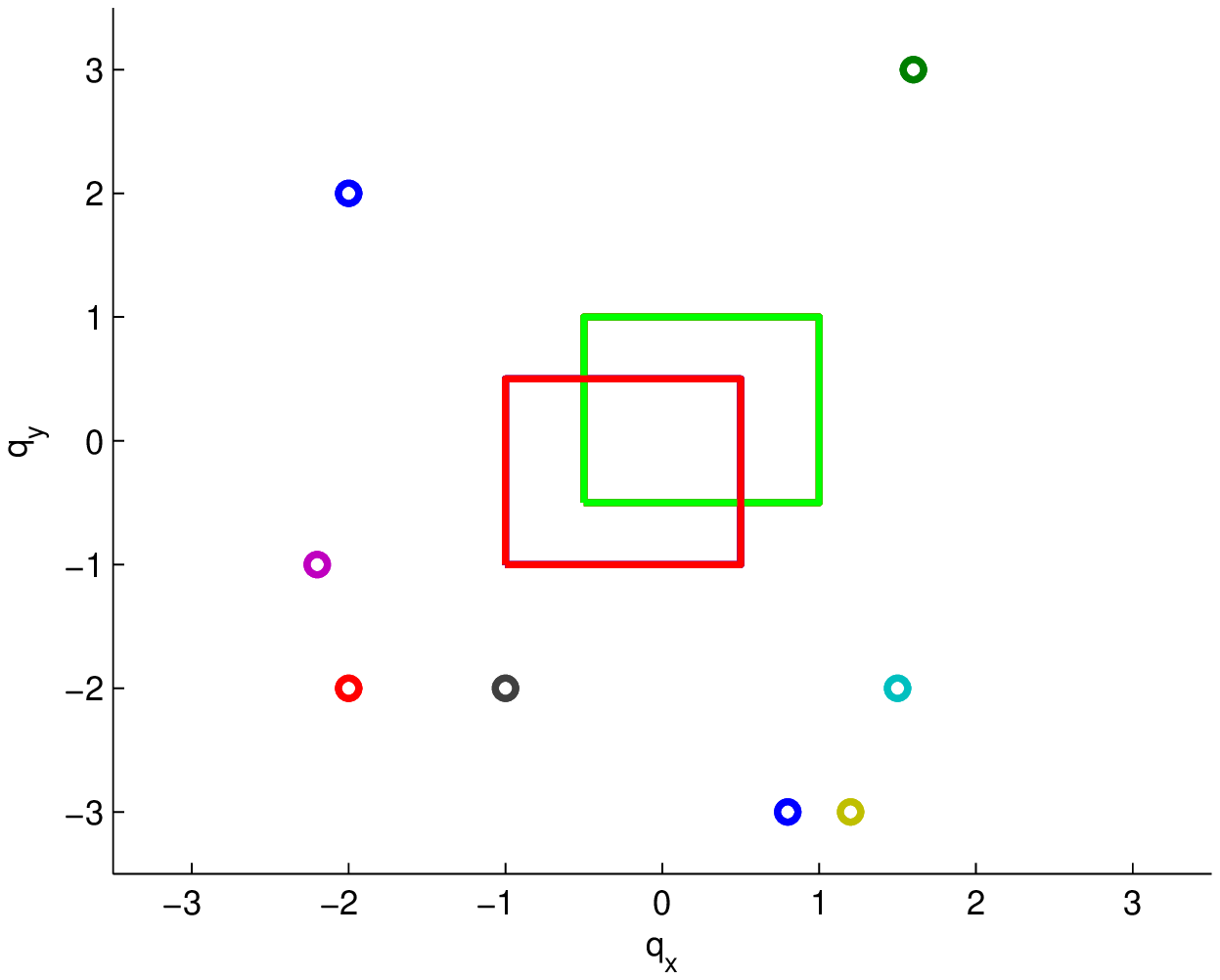}
\includegraphics[scale=0.45]{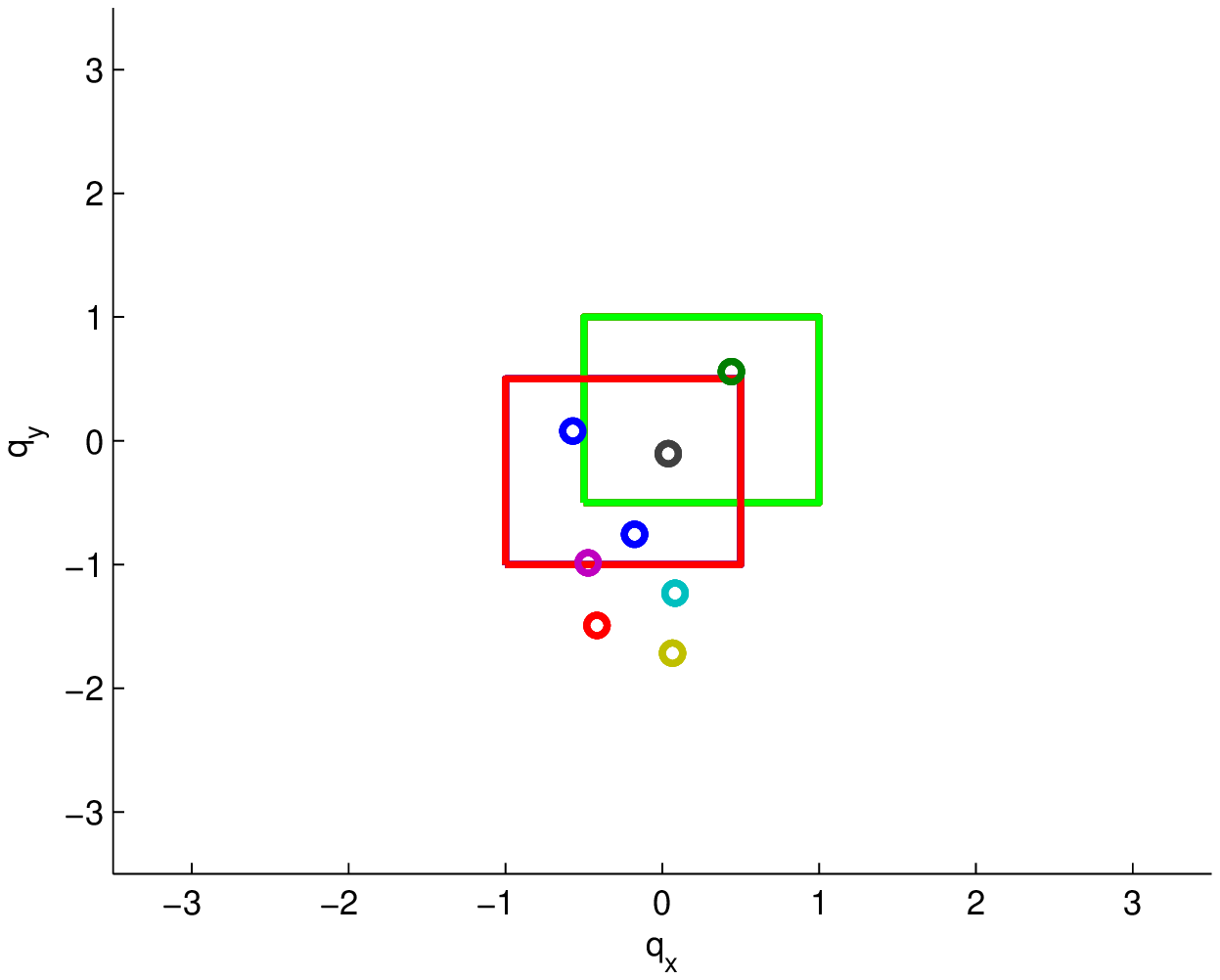}
\includegraphics[scale=0.45]{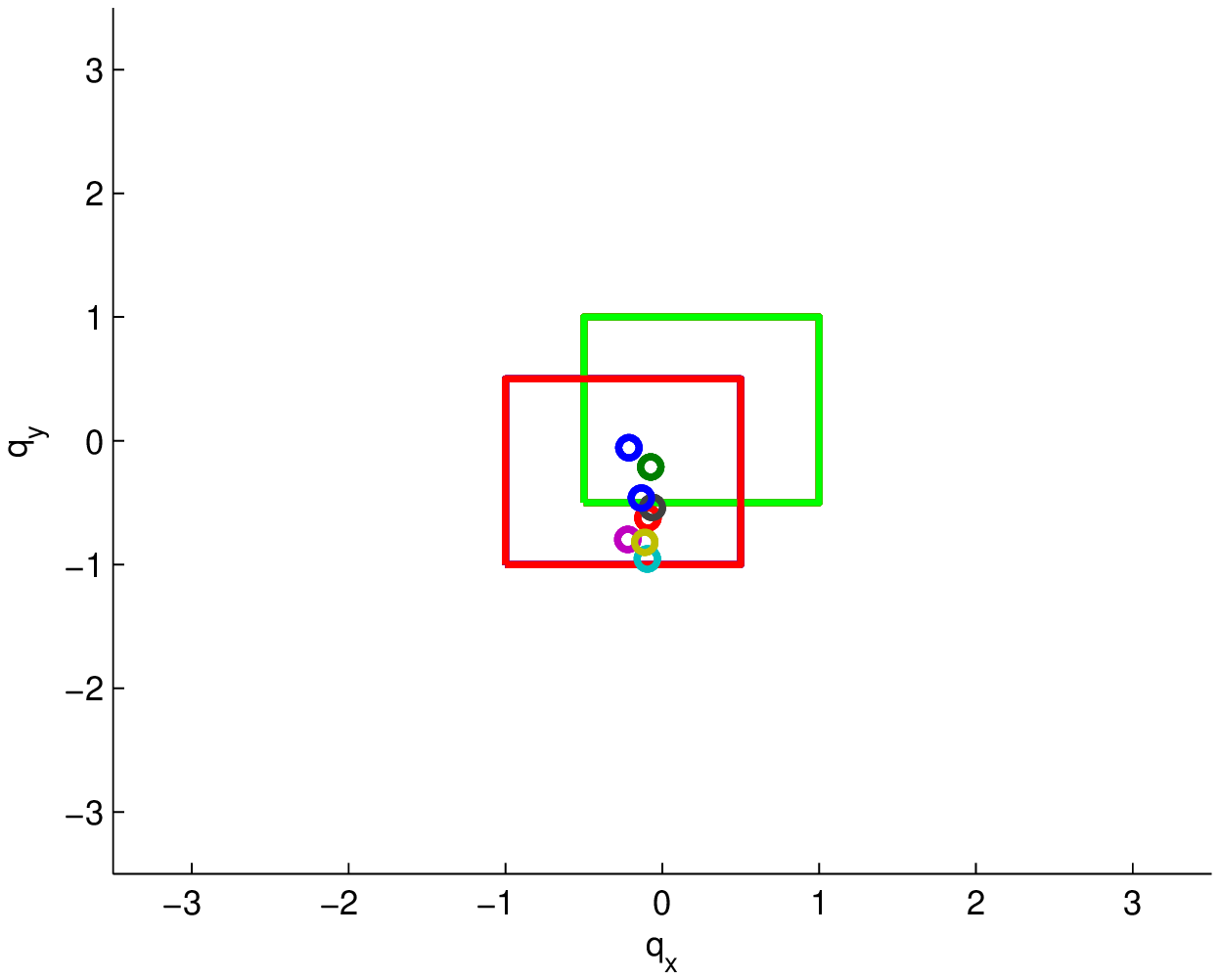}
\includegraphics[scale=0.45]{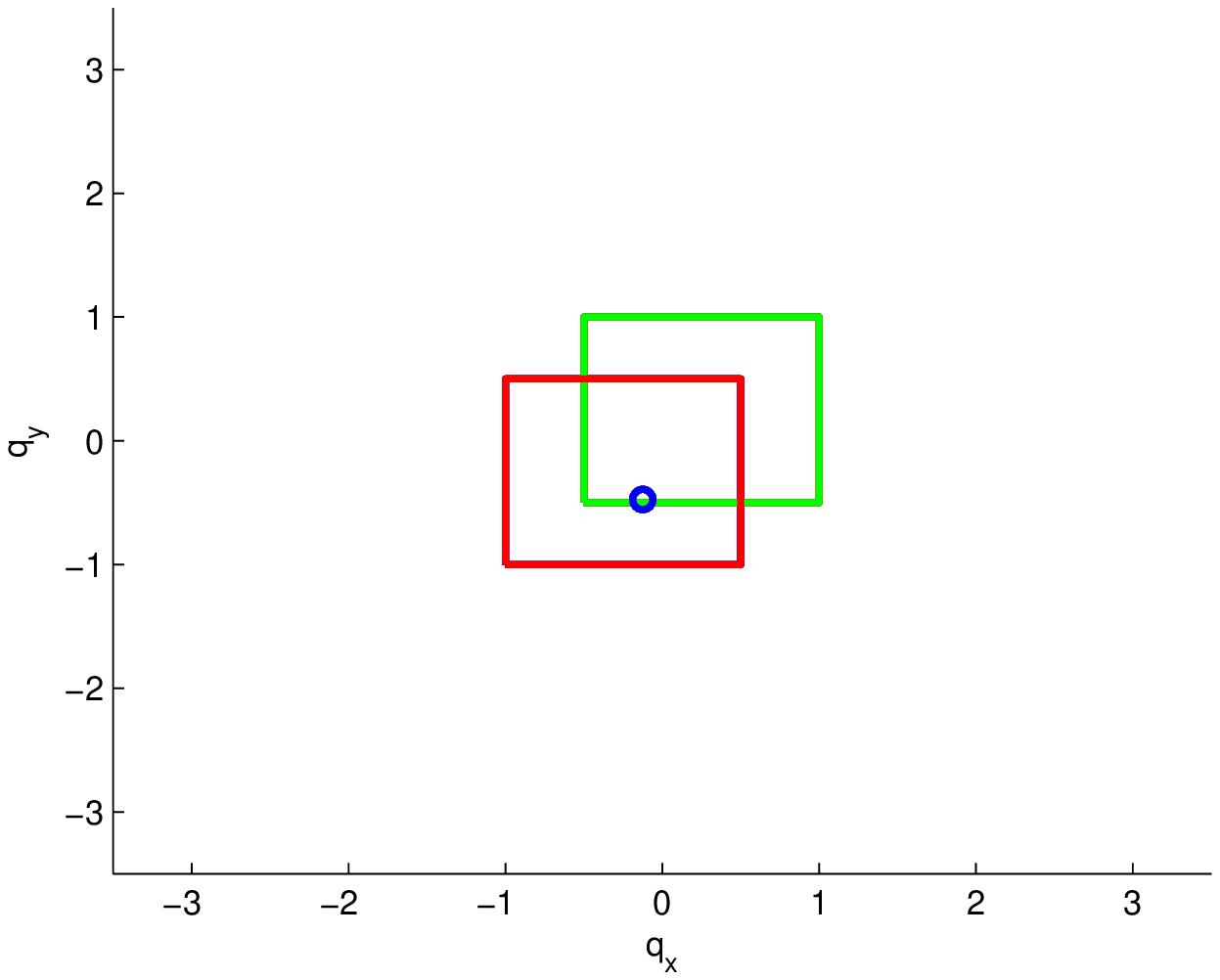}
 \end{center}
\caption{The trajectories of the generalized coordinates of the agents under control \eqref{eq:control2}. The circles denote the
generalized coordinates of the agents.}\label{fig:position3}
\end{figure}

\subsubsection{Switching communication graphs}
\label{sec:simulation2}

we next consider the case of switching communication graphs.
The system dynamics are given the same as those in Section \ref{sec:simulation1}.
The control parameter $k$
is chosen as $k=5$. The available local sets of all the agents are described by rectangles. The shapes of the rectangles are presented  in Fig. \ref{fig:position3} and the initial states of all agents are also the same as those given in Section \ref{sec:simulation1}.
The weight of adjacency matrix $A$ of the generalized coordinates associated with
$\mathcal{G}$ is chosen to be $1$.
The communication graph
$\mathcal{G}$ switches between Fig. \ref{fig:COMMTOP2} and Fig. \ref{fig:COMMTOP3} at time instants $t_{\varrho}=5\varrho$, $\varrho=0,1,\dots$.
\begin{figure}[t]\centering
\begin{tabular} {c}
 \xymatrix{
1\ar@{<->}[r]&  2\ar@{<->}[r] & 3& 4\ar@{<->}[l]\\
8\ar@{<->}[r]           &  7\ar@{<->}[r]           & 6 \ar@{<->}[r]          & 5
} \end{tabular}\caption{The communication graph
$\mathcal{G}$ for Section \ref{sec:simulation2}} \label{fig:COMMTOP2}
\end{figure}
\begin{figure}[t]\centering
\begin{tabular} {c}
 \xymatrix{
1\ar@{<->}[d]&  2 \ar@{<->}[d]& 3\ar@{<->}[d] & 4\ar@{<->}[d]\\
8          &  7          & 6         & 5
} \end{tabular}\caption{The communication graph
$\mathcal{G}$ for Section \ref{sec:simulation2}} \label{fig:COMMTOP3}
\end{figure}

\begin{figure}
 \begin{center}
\includegraphics[scale=0.5]{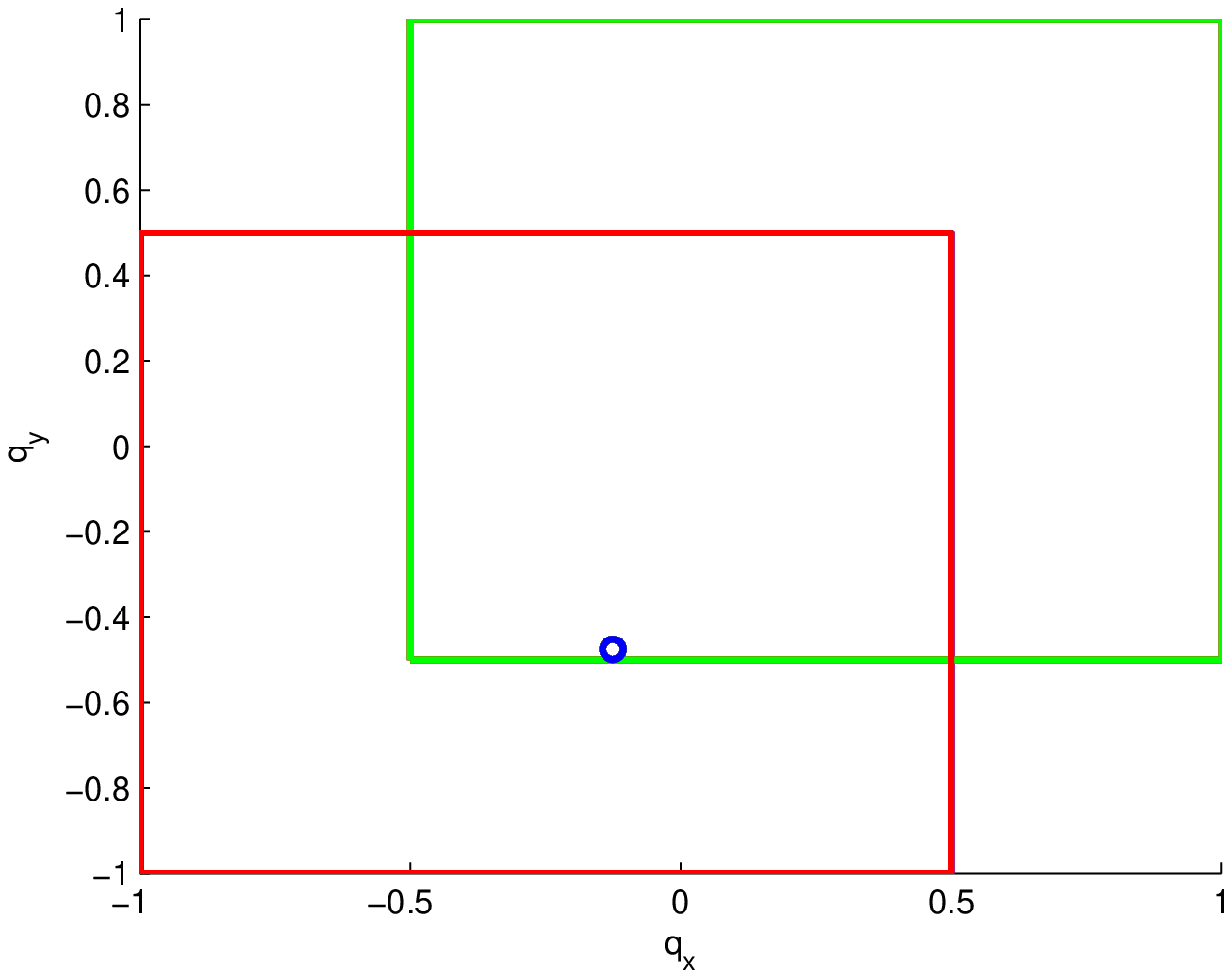}
 \end{center}
\caption{Zoom in of the final generalized coordinates of the agents under control \eqref{eq:control2}.}\label{fig:position3-zoomin}
\end{figure}

\begin{figure}
 \begin{center}
\includegraphics[scale=0.5]{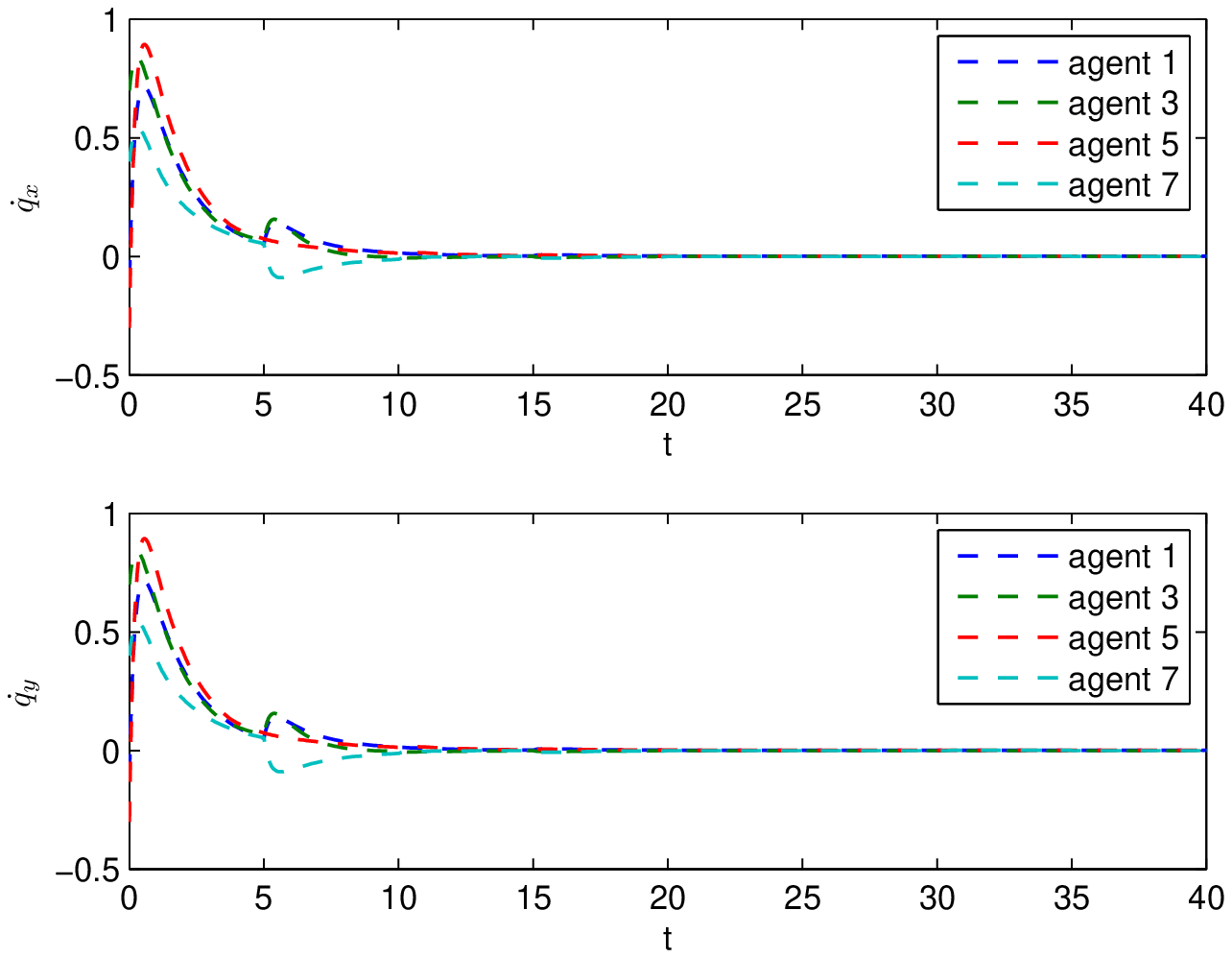}
 \end{center}
\caption{The trajectories of the generalized coordinate derivatives of the agents under control \eqref{eq:control2}.}\label{fig:velocity3}
\end{figure}

Under the control \eqref{eq:control2},
snapshots of generalized coordinates and trajectories of generalized coordinate derivatives of the agents
are shown in
Figs. \ref{fig:position3}, \ref{fig:position3-zoomin} and \ref{fig:velocity3}.
We see that set aggregation is achieved even when the communication graph is switching.

\begin{figure}
 \begin{center}
\includegraphics[scale=0.45]{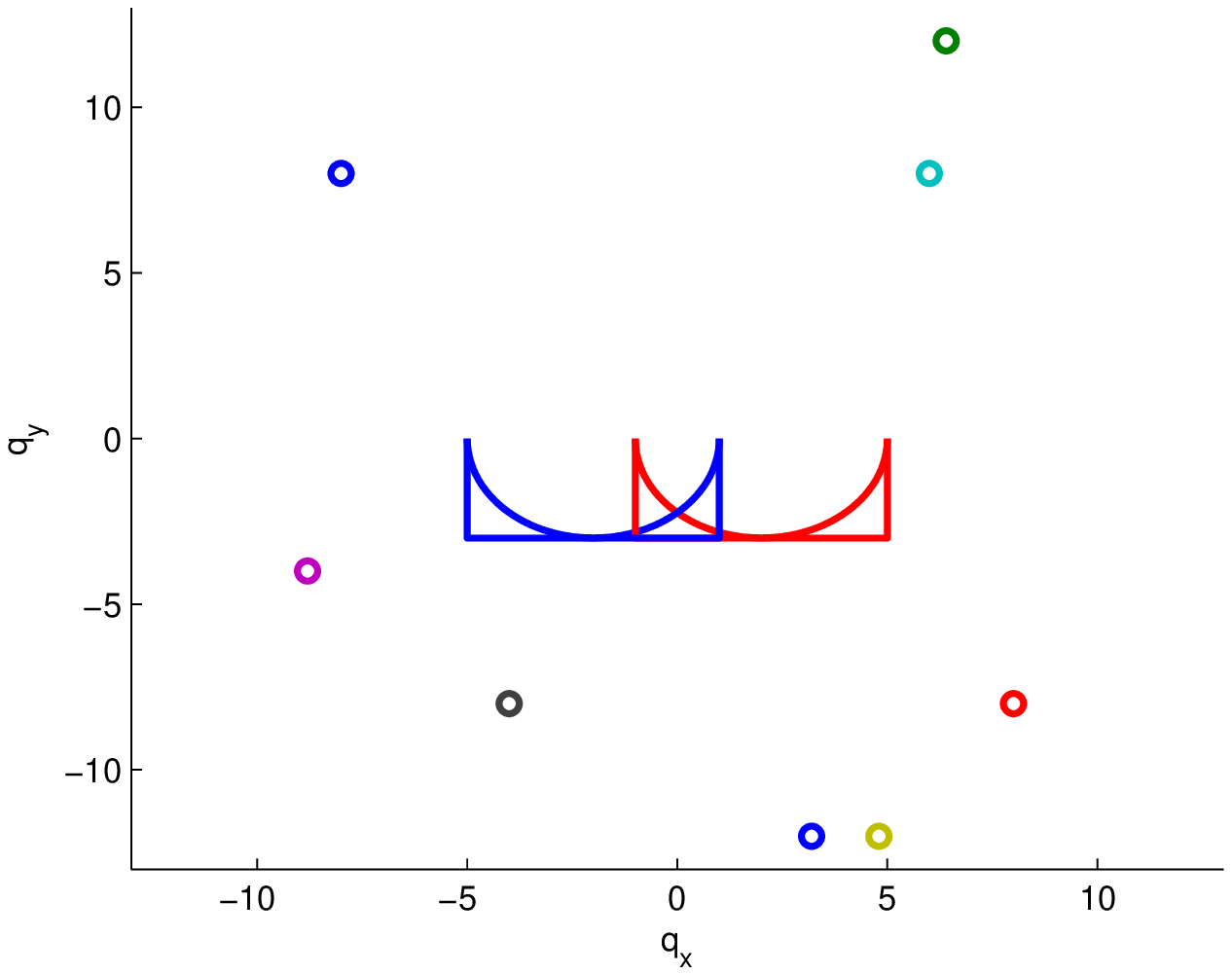}
\includegraphics[scale=0.45]{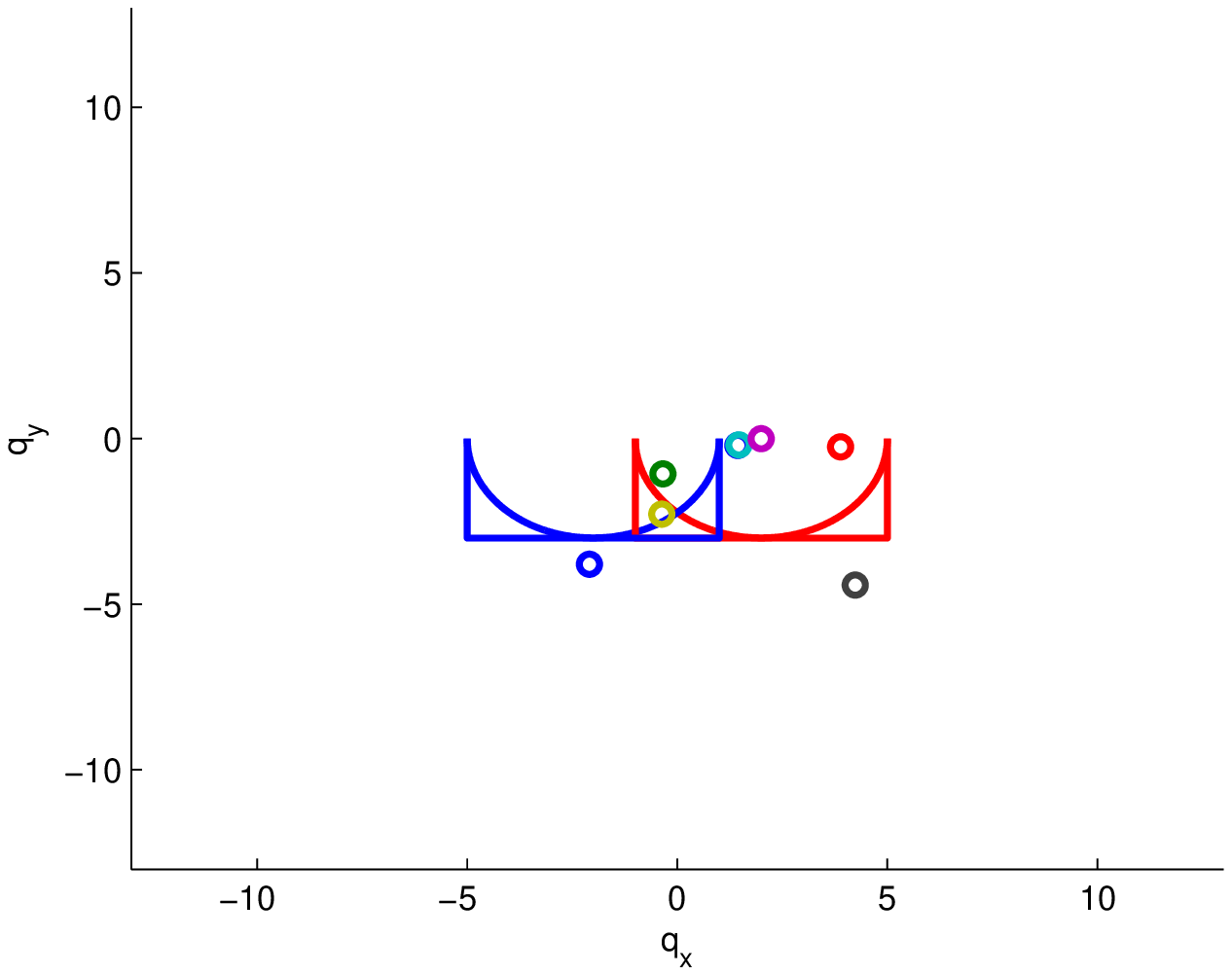}
\includegraphics[scale=0.45]{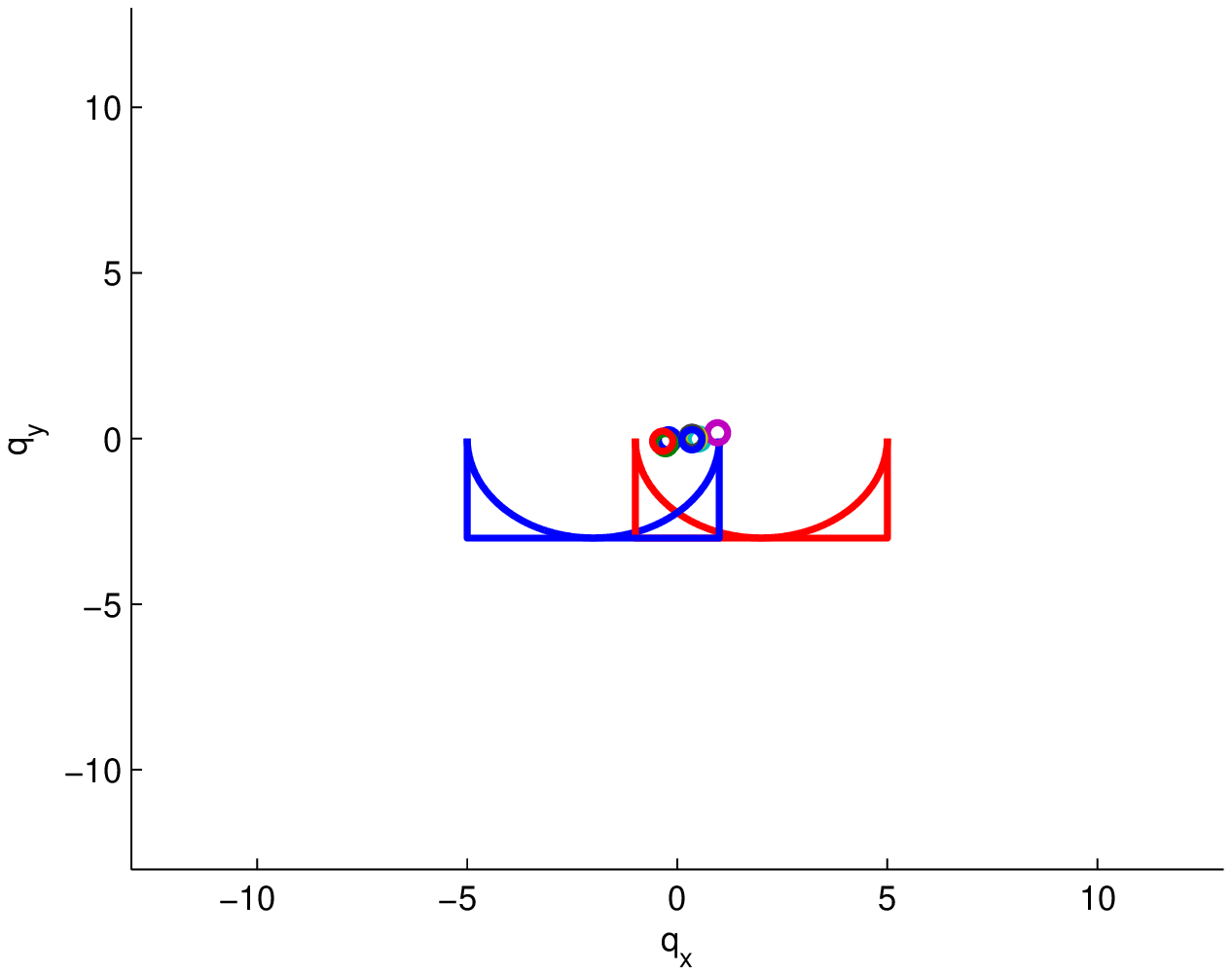}
\includegraphics[scale=0.45]{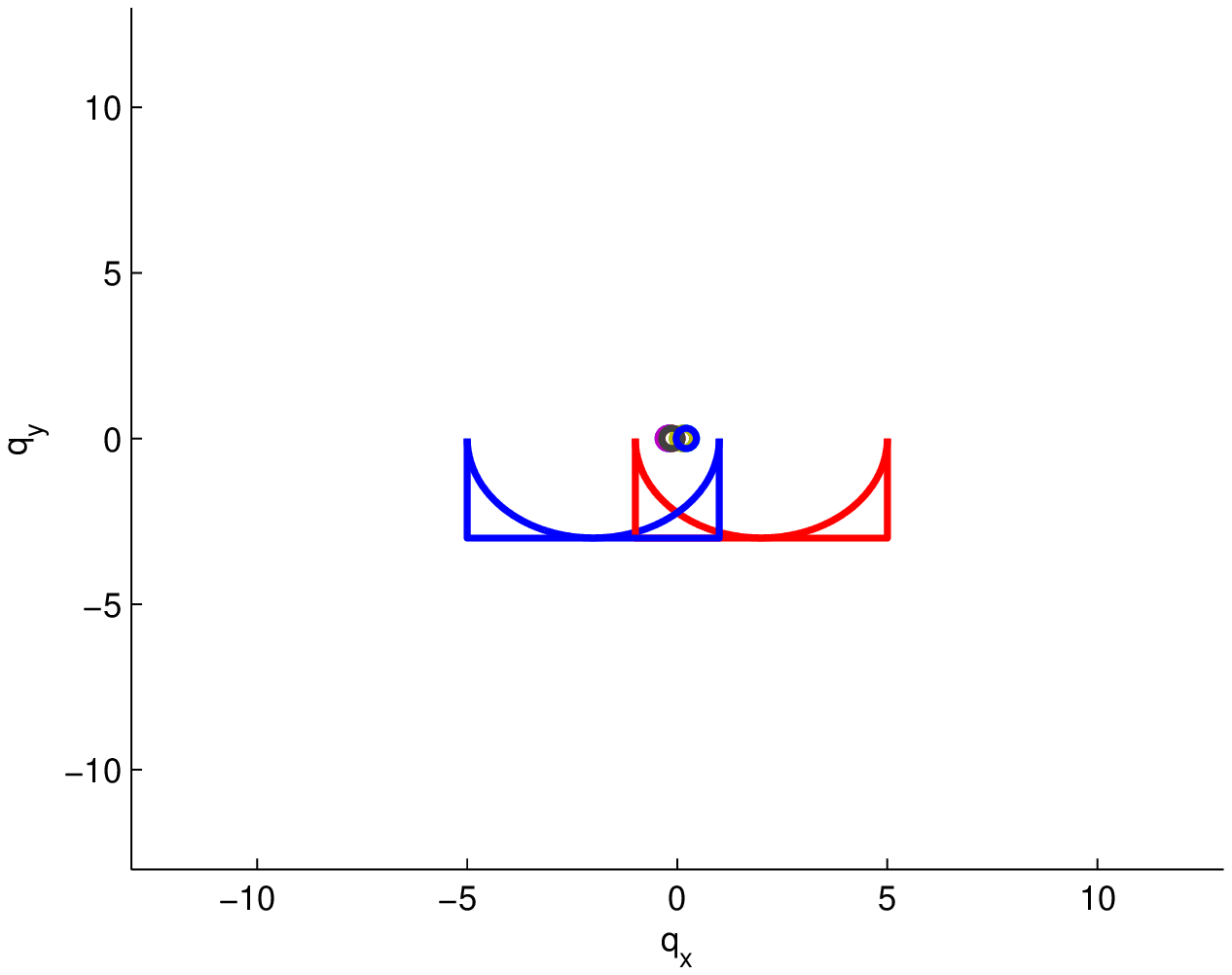}
 \end{center}
\caption{The trajectories of the generalized coordinates of the agents under control \eqref{eq:control1}. The circles denote the
generalized coordinates of the agents.}\label{fig:position2}
\end{figure}

\subsubsection{Discussions on the case of non-convex local sets}

we next consider the case where $\mathcal{X}_i$, for all $i\in\mathcal{V}$ are non-convex sets. We assume that the available local sets of all the agents have irregular forms as shown in the Fig. \ref{fig:position2}.
The initial states of the agents, the control parameter, the communication graph
$\mathcal{G}$ are the same as those for Section \ref{sec:simulation1}.

\begin{figure}
 \begin{center}
\includegraphics[scale=0.5]{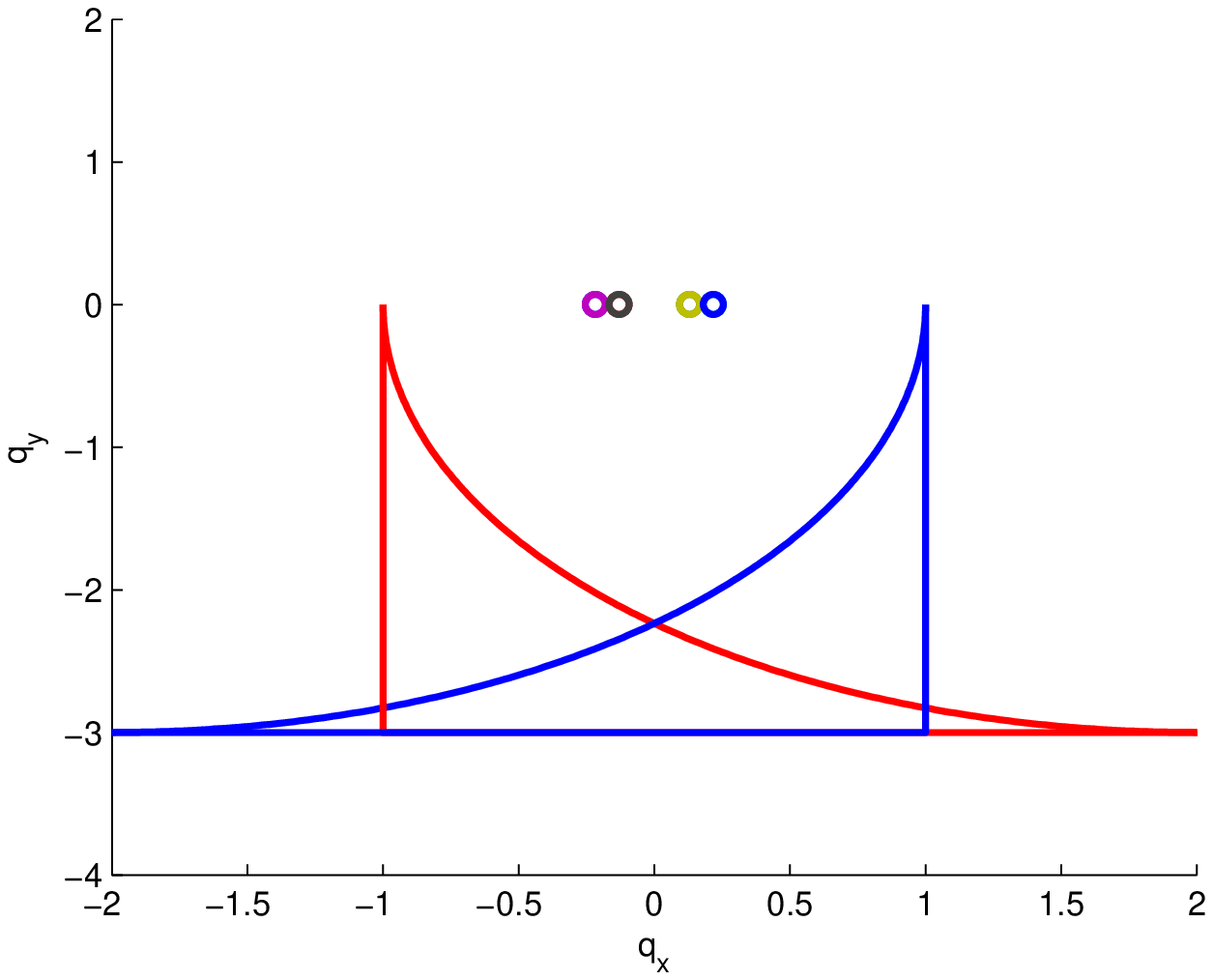}
 \end{center}
\caption{Zoom in of the final generalized coordinates of the agents under control \eqref{eq:control1}}\label{fig:position2-zoomin}
\end{figure}

\begin{figure}
 \begin{center}
\includegraphics[scale=0.5]{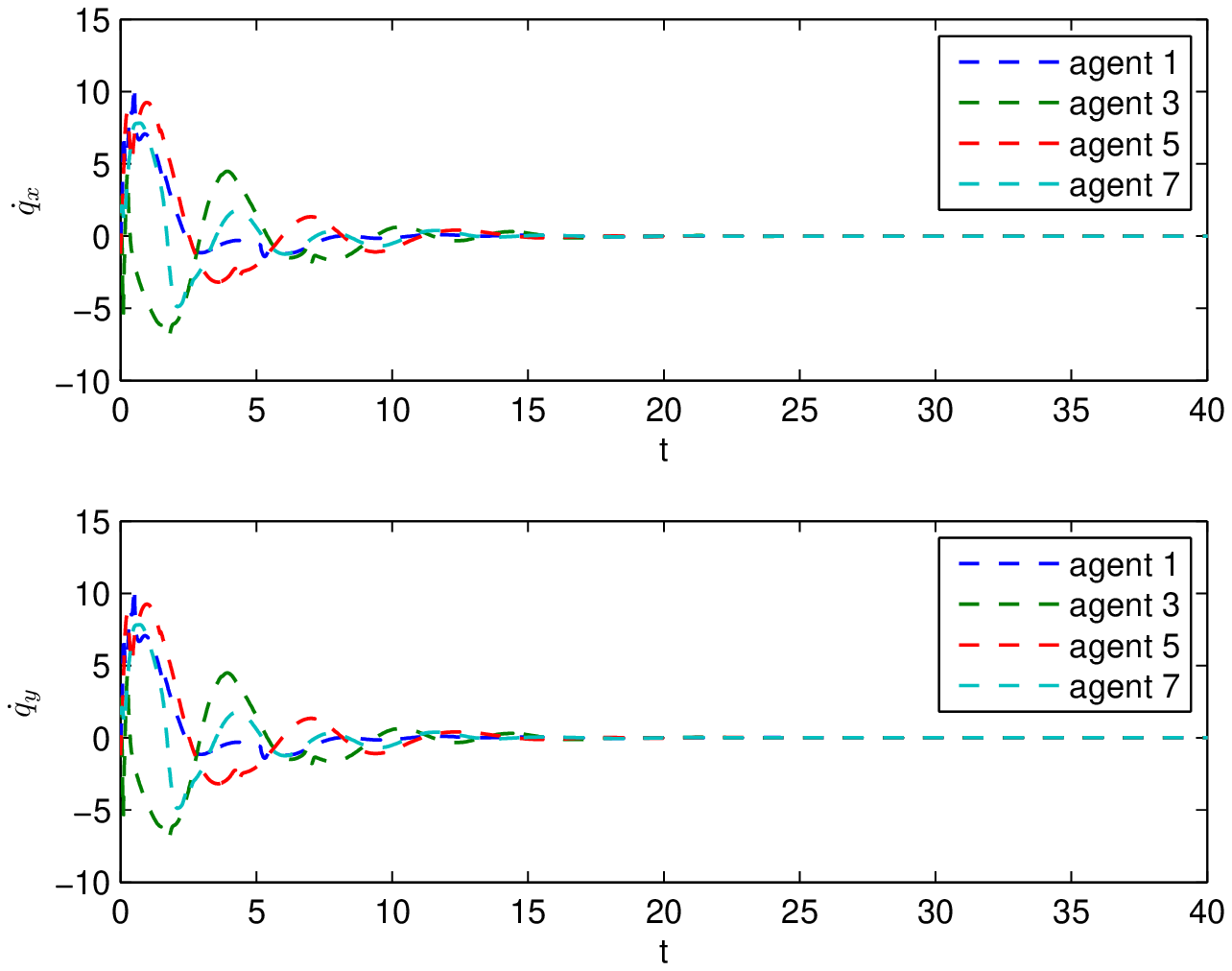}
 \end{center}
\caption{The trajectories of the generalized coordinate derivatives of the agents under control \eqref{eq:control1}.}\label{fig:velocity2}
\end{figure}

Under the control \eqref{eq:control1},
snapshots of generalized coordinates and trajectories of generalized coordinate derivatives of the agents
are shown in
Figs. \ref{fig:position2}, \ref{fig:position2-zoomin} and \ref{fig:velocity2}.
We see that set aggregation cannot be achieved. All the agents neither converge into their local sets nor reach a consensus.
One way to overcome this side effect is to construct the convex hulls of all the local non-convex sets. By projecting the agents' generalized coordinates onto the convex hulls of the local sets, we can still guarantee the set aggregation on the intersection of the convex hulls of the local sets using the proposed control. All these observations reveal the fundamental importance of the convexity of local sets in order for cooperative set aggregation.


\section{Cooperative aggregation with collision avoidance}\label{sec:collision}

In certain practical applications, in addition to track a common objective using the constrained local information and information exchange, all the agents also need to guarantee collision avoidance during the movement. In such a case, global set aggregation cannot be achieved since there exist minimum safety distances between each pair of agents. Instead, multiple Lagrangian systems approach a bounded region near the intersection of all the local target sets while the collision avoidance is ensured during the movement. In this section, we investigate the considered cooperative set aggregation problem with collision avoidance also taken into consideration.

\subsection{Fixed Communication Graph: Approximate Aggregation}

\begin{figure}
 \begin{center}
\includegraphics[scale=0.3]{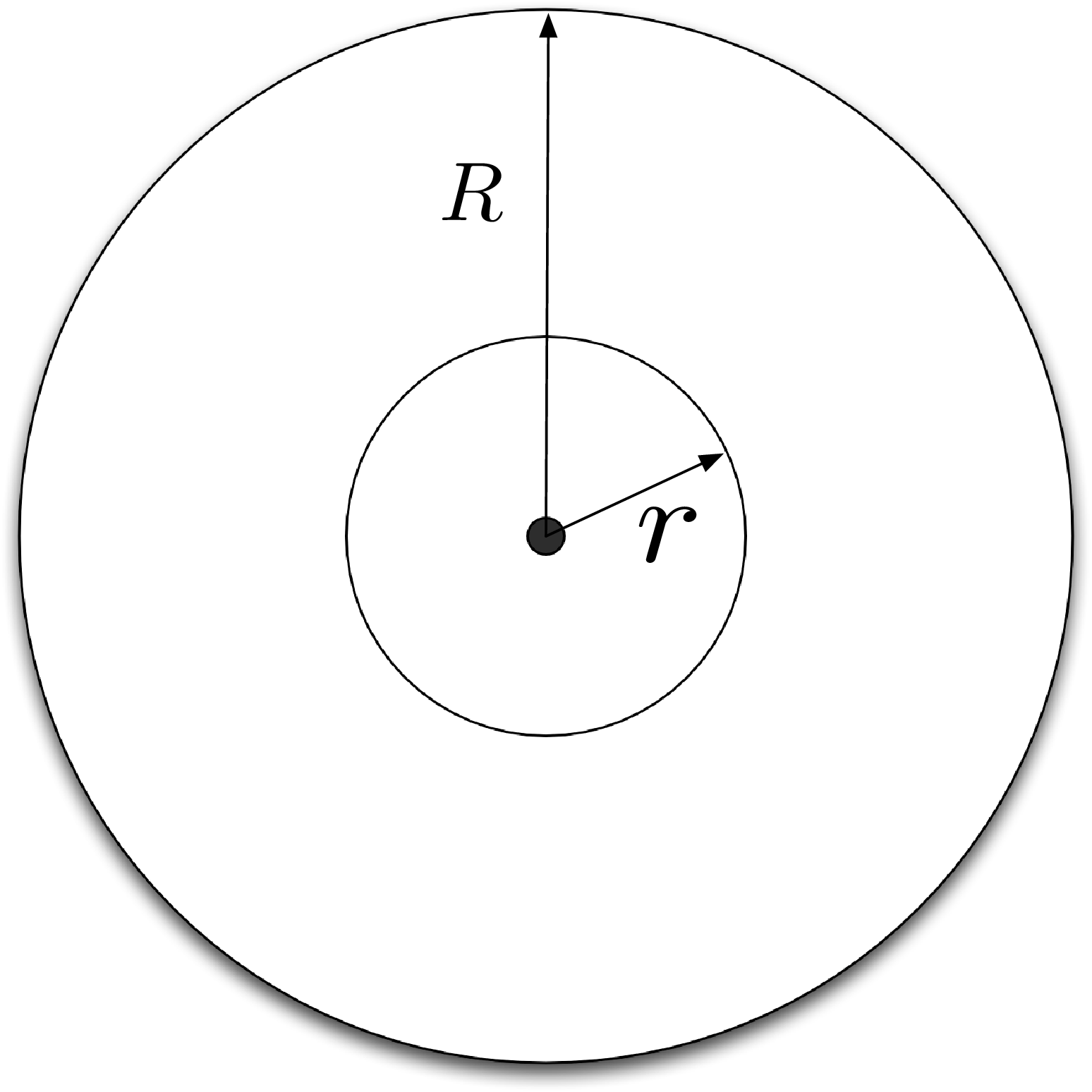}
 \end{center}
\caption{Sensing radius $R$ and minimum safety radius $r$}\label{fig:avoidance}
\end{figure}

In this section, we assume that the communication graph is fixed. All the agents are also equipped with inter-agent sensors, where the sensing radius is denoted by $R$ and minimum safety radius is denoted by $r$ (see Fig. \ref{fig:avoidance} for the illustration).

The following cooperative control with collision avoidance is proposed for all $i\in\mathcal{V}$,
\begin{equation}
\tau_i=\underbrace{-k\dot q_i}_{\tau_i^v}\underbrace{-(q_i-P_{\mathcal{X}_i}(q_i))}_{\tau_i^{s}}\underbrace{-\sum_{j\in \mathcal{N}_i}a_{ij}(q_i-q_j)}_{\tau_i^{in}}\underbrace{-\sum_{j=1}^n\frac{\partial V_{ij}}{\partial q_i}}_{\tau_i^{a}}, \label{eq:control3}
\end{equation}
where $k>0$ denotes generalized coordinate derivative damping, and motivated by \cite{Spong_AMC10}, we define $V_{ij}=V_{ij}(\|q_i-q_j\|)$ as a positive function:
\begin{equation*}
V_{ij}=\begin{cases}
 0, \quad ~{\rm if}~ \|q_i-q_j\|\geq R  \\
   \left(\frac{\|q_i-q_j\|-R^2}{\|q_i-q_j\|-r^2}\right)^2, ~~{\rm if}~~ r < \|q_i-q_j\|< R,
\end{cases}
\end{equation*}
where $R>r>0$ are positive constants. Then, we have that
\begin{equation*}
\frac{\partial V_{ij}}{\partial q_i} =
\begin{cases}
 0, \quad ~\|q_i-q_j\|\geq R  \\
    \frac{4(R^2-r^2)(\|q_i-q_j\|^2-R^2)}{(\|q_i-q_j\|^2-r^2)^3}(q_i-q_j),   r < \|q_i-q_j\|< R.
\end{cases}
\end{equation*}
Note that $V_{ij}$ is a differentiable, nonnegative function of the relative distance $\|q_i-q_j\|$ defined in $(r,\infty)$. Also note that $V_{ij}(\|q_i-q_j\|)\rightarrow\infty$ as $\|q_i-q_j\|\rightarrow r^+$.

Let $\mathcal{S}_1,\mathcal{S}_2,\dots,\mathcal{S}_r$ be closed convex sets with intersection $\mathcal{S}_0=\bigcap_{i=1}^r\mathcal{S}_i$ is nonempty. The collection $\{ \mathcal{S}_1,\mathcal{S}_2,\dots,\mathcal{S}_r\}$ is said to be linearly regular \cite{Frank_2008} if there exists $\rho>0$ such that $\|x\|_{\mathcal{S}_0}\leq \rho \max_{i=1,2,\dots,r}\|x\|_{\mathcal{S}_i}$ for all $x\in \mathbb{R}^m$.

The main result we obtain is as follows, and it turns out certain approximate cooperative set aggregation can be achieved with collision avoidance guarantee.
\begin{them}
Let $\{\mathcal{X}_1,\mathcal{X}_2,\dots,\mathcal{X}_n\}$ be linearly regular and $\|q_i(t_0)-q_j(t_0)\|>r$, for all $i,j\in\mathcal{V}$. Suppose that Assumption 1 holds and the communication graph $\mathcal{G}$ is connected. Then the closed-loop system \eqref{eq:Lagrangian} with \eqref{eq:control3} ensures
\begin{enumerate}
\item $\|q_i(t)-q_j(t)\|>r,\quad \forall i,j\in\mathcal{V}, \quad \forall t\geq t_0\geq 0$, where $r$ is a positive constant,
\item $\lim_{t\rightarrow \infty}\|q_i(t)\|_{\mathcal{X}_0}\leq B^*,\quad \forall i\in\mathcal{V}$, where $B^*$ is a positive constant,
\item $\lim_{t\rightarrow \infty}\dot q_i(t)=0,\quad \forall i\in\mathcal{V}$.
\end{enumerate}
\end{them}
\proof
We propose the following Lyapunov function
\begin{align}
V=&\frac{1}{2}\sum_{i=1}^n\dot q_i\T M_i(q_i) \dot q_i+\frac{1}{4}\sum_{i=1}^n\sum_{j\in \mathcal{N}_i}a_{ij}\|q_i-q_j\|^2
\notag\\&+\frac{1}{2}\sum_{i=1}^n\|q_i-P_{\mathcal{X}_i}(q_i)\|^2+\frac{1}{2}\sum_{i=1}^n\sum_{j=1}^nV_{ij}
\label{eq:V3}
\end{align}
Note that the control term $\frac{\partial V_{ij}}{\partial q_i}$ does not introduce discontinuities because the potential function
is differentiable at the transition point.
The derivative of $V$ along \eqref{eq:Lagrangian} with \eqref{eq:control3} is
\begin{align*}
\dot V=&~\sum_{i=1}^n\dot q_i\T \left(-k\dot q_i-\sum_{j\in \mathcal{N}_i}a_{ij}(q_i-q_j)-(q_i-P_{\mathcal{X}_i}(q_i))\right.
\\&\left.-\sum_{j=1}^n\frac{\partial V_{ij}}{\partial q_i}\right)
+\frac{1}{2}\sum_{i=1}^n\sum_{j\in \mathcal{N}_i}a_{ij}(q_i-q_j)\T (\dot q_i-\dot q_j)
\\&+\sum_{i=1}^n\dot q_i\T(q_i-P_{\mathcal{X}_i}(q_i))+\frac{1}{2}\sum_{i=1}^n (\dot q_i-\dot q_j)\T\sum_{j=1}^n\frac{\partial V_{ij}}{\partial q_i}
\\=&~-k\sum_{i=1}^n\dot q_i\T \dot q_i-\sum_{i=1}^n\dot q_i\T\sum_{j\in \mathcal{N}_i}a_{ij}(q_i-q_j)
\\&+\sum_{i=1}^n\dot q_i\T\sum_{j\in \mathcal{N}_i}a_{ij}(q_i-q_j)
\\=&~-k\sum_{i=1}^n\|\dot q_i\|^2
\\ \leq &~0,
\end{align*}
where we have used the fact $\frac{\partial V_{ij}}{\partial (q_i-q_j)}=\frac{\partial V_{ij}}{\partial q_i}=-\frac{\partial V_{ij}}{\partial q_j}=\frac{\partial V_{ji}}{\partial q_i}=-\frac{\partial V_{ij}}{\partial q_j}$, for all $i,j\in\mathcal{V}$. Therefore, we know that $V(t)\leq V(t_0)<\infty$ is bounded for all $t\geq t_0$. This further implies that $V_{ij}(t)\leq V(t_0)$ is bounded for all $i,j\in \mathcal{V}$ and for all $t\geq t_0$. This shows that the collision avoidance is guaranteed, {\em i.e.,} $\|q_i(t)-q_j(t)\|>r,~\forall i,j\in\mathcal{V},~ \forall t\geq t_0\geq 0$ since $V_{ij}(\|q_i-q_j\|)\rightarrow\infty$ as $\|q_i-q_j\|\rightarrow r^+$.

Also, based on LaSalle's Invariance Principle (Theorem 4.4 of \cite{Khalil_book}), we know that every solution of \eqref{eq:Lagrangian} with \eqref{eq:control3} converges to largest invariant set in $\mathcal{M}$, where $\mathcal{M}=\{q_i\in \mathbb{R}^m,\dot q_i\in \mathbb{R}^m,~\forall i\in \mathcal{V} ~|~ \dot q_i=0,\forall i\in \mathcal{V}\}$. Let $q_i(t)$, $\dot q_i(t)$, $\forall i\in \mathcal{V}$ be a solution that belongs to $\mathcal{M}$:
\begin{align*}
\dot q_i\equiv 0,~~\forall i\in \mathcal{V} \Rightarrow \ddot q_i\equiv 0,~~\forall i\in \mathcal{V}.
\end{align*}
It thus follows that for all $i\in \mathcal{V}$,
\begin{align*}
-\sum_{j\in \mathcal{N}_i}a_{ij}(q_i-q_j)-(q_i-P_{\mathcal{X}_i}(q_i))-\sum_{j=1}^n\frac{\partial V_{ij}}{\partial q_i}\equiv 0.
\end{align*}
It also follows from \eqref{eq:V3} that $\|q_i(t)\|_{\mathcal{X}_i}\leq \sqrt{2V(t_0)}$, for all $i\in\mathcal{V}$, for all $t\geq t_0$ and that $q\T(t)(L\otimes I_m)q(t)\leq 2V(t_0)$, for all $t\geq t_0$, where $q=[q_1\T,q_2\T,\dots,q_n\T]$. Since the graph is undirected connected, we can sort the eigenvalues of Laplacian matrix $L\otimes I_m$ as
\begin{align*}
0=\lambda_1=\dots=\lambda_m<\lambda_{m+1}=\dots=\lambda_{2m}\leq\dots\leq \lambda_{mn}.
\end{align*}
Let $\phi_1,\phi_2,\dots,\phi_{mn}$ be the orthonormal basis of $\mathbb{R}^{mn}$ formed by the right eigenvectors of $L\otimes I_m$, where $\phi_1,\dots,\phi_m$ are the eigenvectors corresponding to the zero eigenvalue. Suppose $q=\sum_{k=1}^{mn}c_k\phi_k$ with $c_k\in\mathbb{R}$, $k=1,2,\dots,mn$. It then follows that
\begin{align*}
q\T(t)(L\otimes I_m)q(t)=\sum_{k=m+1}^{mn}c_k^2\phi_k^2\lambda_k=\sum_{k=m+1}^{mn}c_k^2\lambda_k\leq 2V(t_0).
\end{align*}
Therefore, we know that
\begin{align*}
\sum_{k=m+1}^{mn}c_k^2\leq \frac{2V(t_0)}{\lambda_{m+1}},
\end{align*}
where $\lambda_{m+1}$ denotes the smallest non-zero eigenvalue of $L$. Define the consensus manifold as $\mathcal{U}=\{q\in \mathbb{R}^{mn}:q_1=q_2\dots=q_n\}$. Since $\mathcal{U}={\rm span}\{\phi_1,\phi_2,\dots,\phi_{mn}\}$, we know that
\begin{align*}
\sum_{k=m+1}^{mn}c_k^2=\|q\|^2_{\mathcal{U}}=\sum_{i=1}^n\|q_i-q_{c}\|^2\leq \frac{2V(t_0)}{\lambda_{m+1}},
\end{align*}
where $q_c=\frac{\sum_{i=1}^nq_i}{n}$. Therefore, we can conclude that for all $i,j\in\mathcal{V}$,
\begin{align*}
\|q_i-q_j\|\leq 2 \sqrt{\frac{V(t_0)}{\lambda_{m+1}}}.
\end{align*}
Therefore, we know that
$\|q_i(t)\|_{\mathcal{X}_k}\leq (\sqrt{2}+\frac{2}{\sqrt{\lambda_{m+1}}})\sqrt{V(t_0)}$, for all $i,k\in\mathcal{V}$, for all $t\geq t_0$.
Then, based on the definition of linearly regularity, we know that $\|q_i(t)\|_{\mathcal{X}_0}\leq \rho (\sqrt{2}+\frac{2}{\sqrt{\lambda_{m+1}}})\sqrt{V(t_0)}$, for all $i\in\mathcal{V}$, where $\rho$ is a known constant.
Therefore, the desired conclusion follows.
\endproof

\subsection{Simulation Verifications}

We now use numerical simulations to validate the effectiveness of the theoretical results obtained in Section \ref{sec:collision}. We assume that there are sixteen agents ($n=16$) in the group.
For the case of set aggregation with collision avoidance, the control parameter $k$
is chosen as $k=1$, the sensing radius is given by $R=2$ and the minimum safety distance is $r=0.2$. The local available sets are considered rectangles and the shapes of the rectangles are given in Section \ref{sec:simulation2}.

\subsubsection{Star graph}
we first consider the case when the communication graph is a star graph. Agent $1$ is assumed to be the center of the star and the communication weight is chosen to be $1$.

\begin{figure}
 \begin{center}
\includegraphics[scale=0.45]{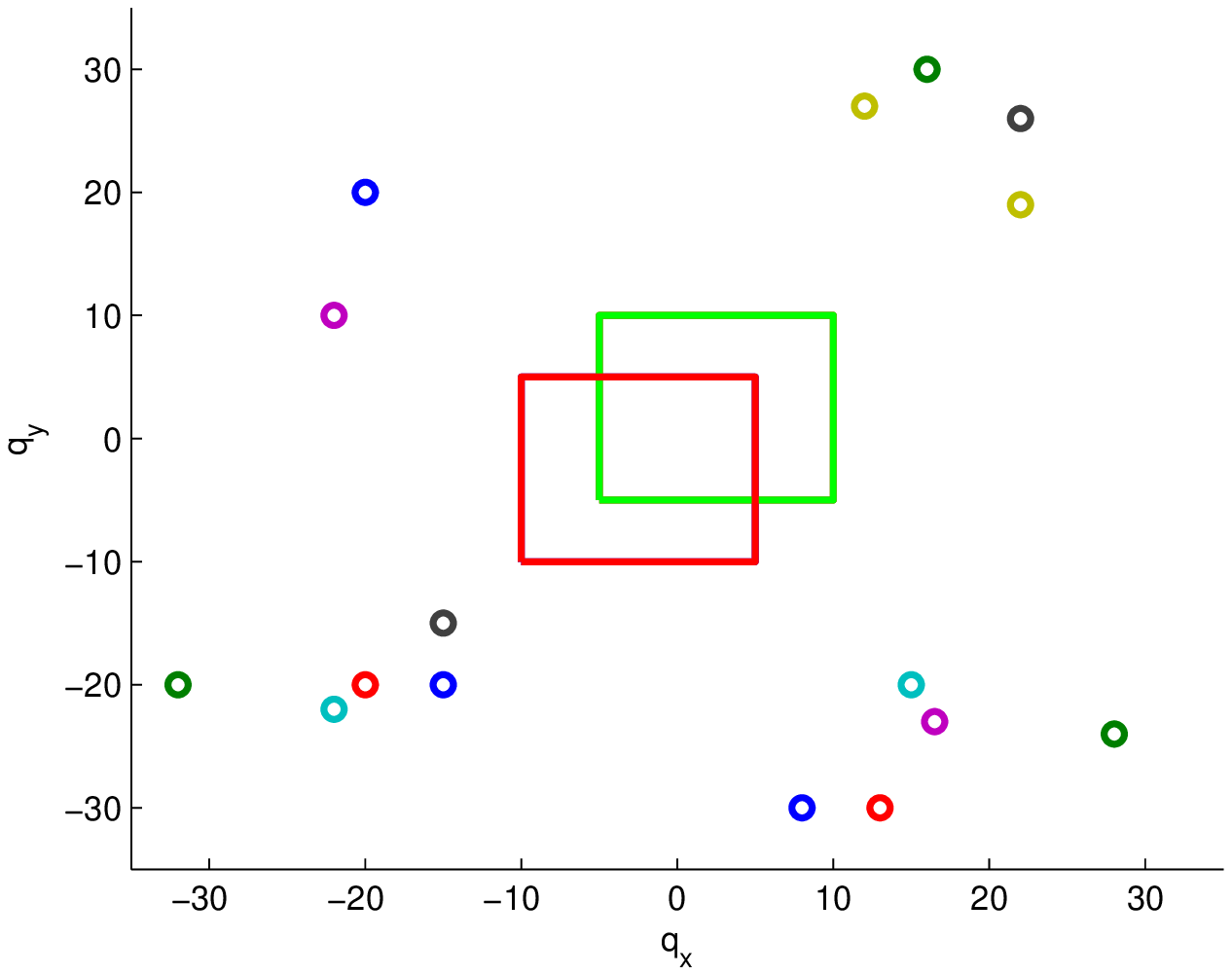}
\includegraphics[scale=0.45]{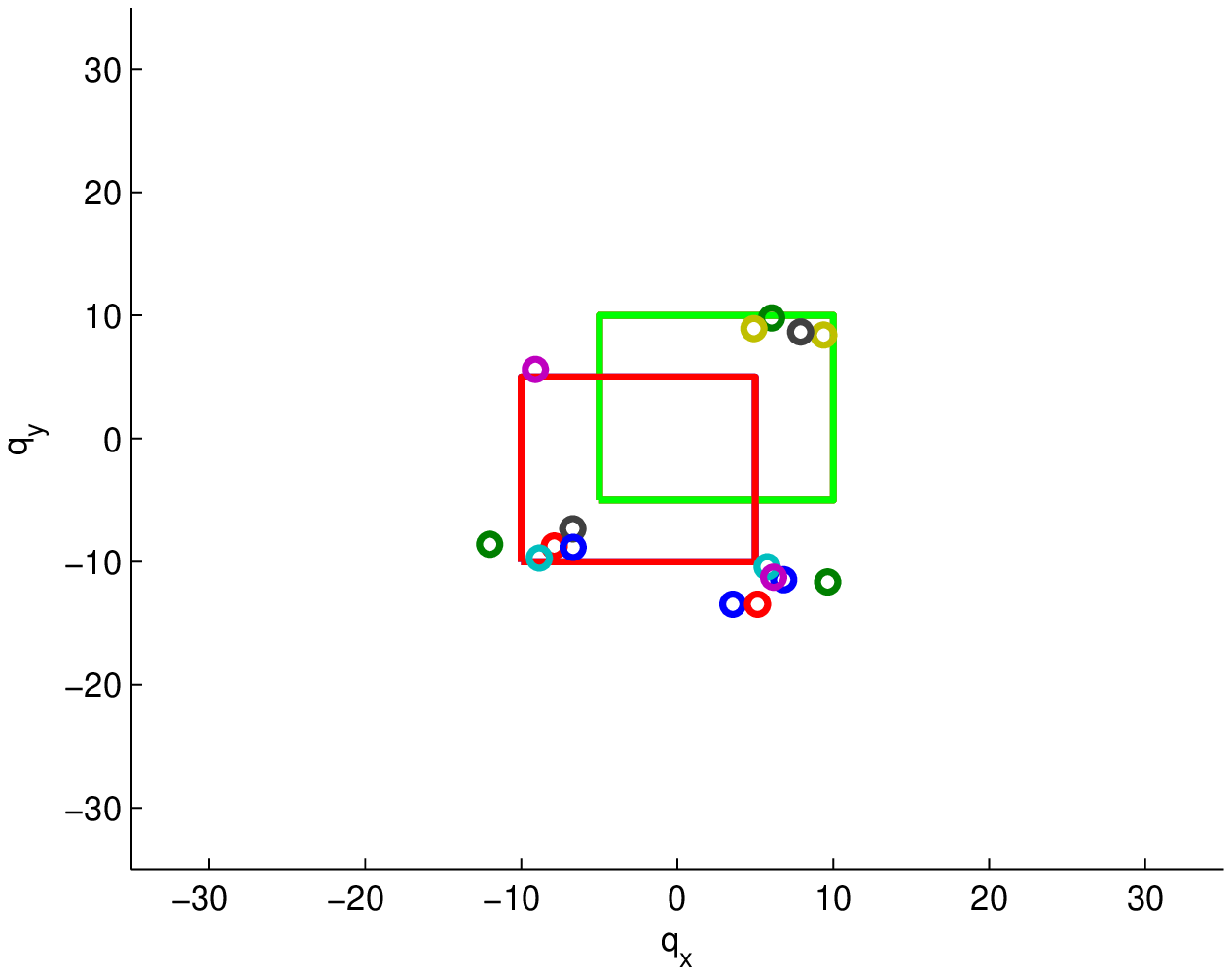}
\includegraphics[scale=0.45]{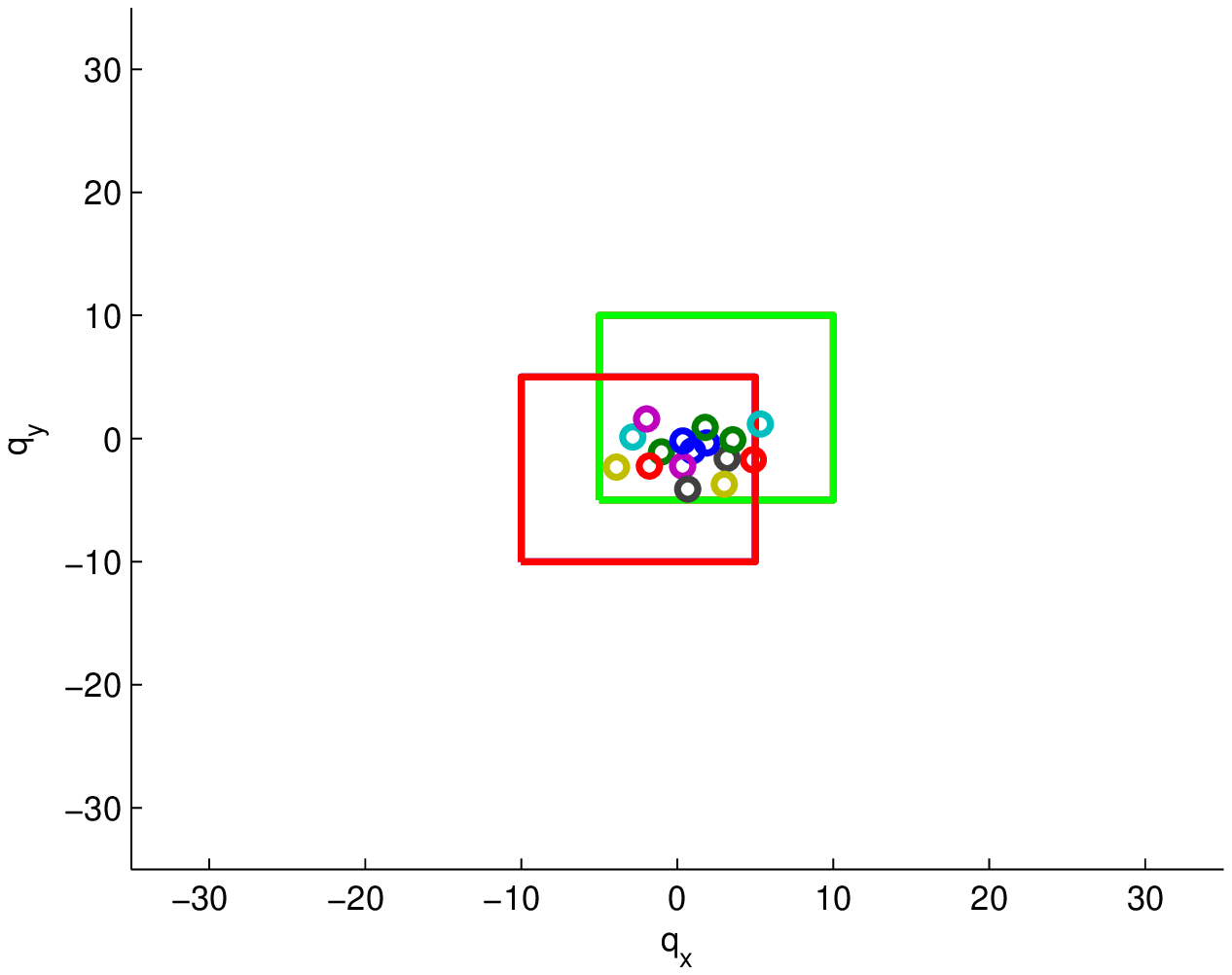}
\includegraphics[scale=0.45]{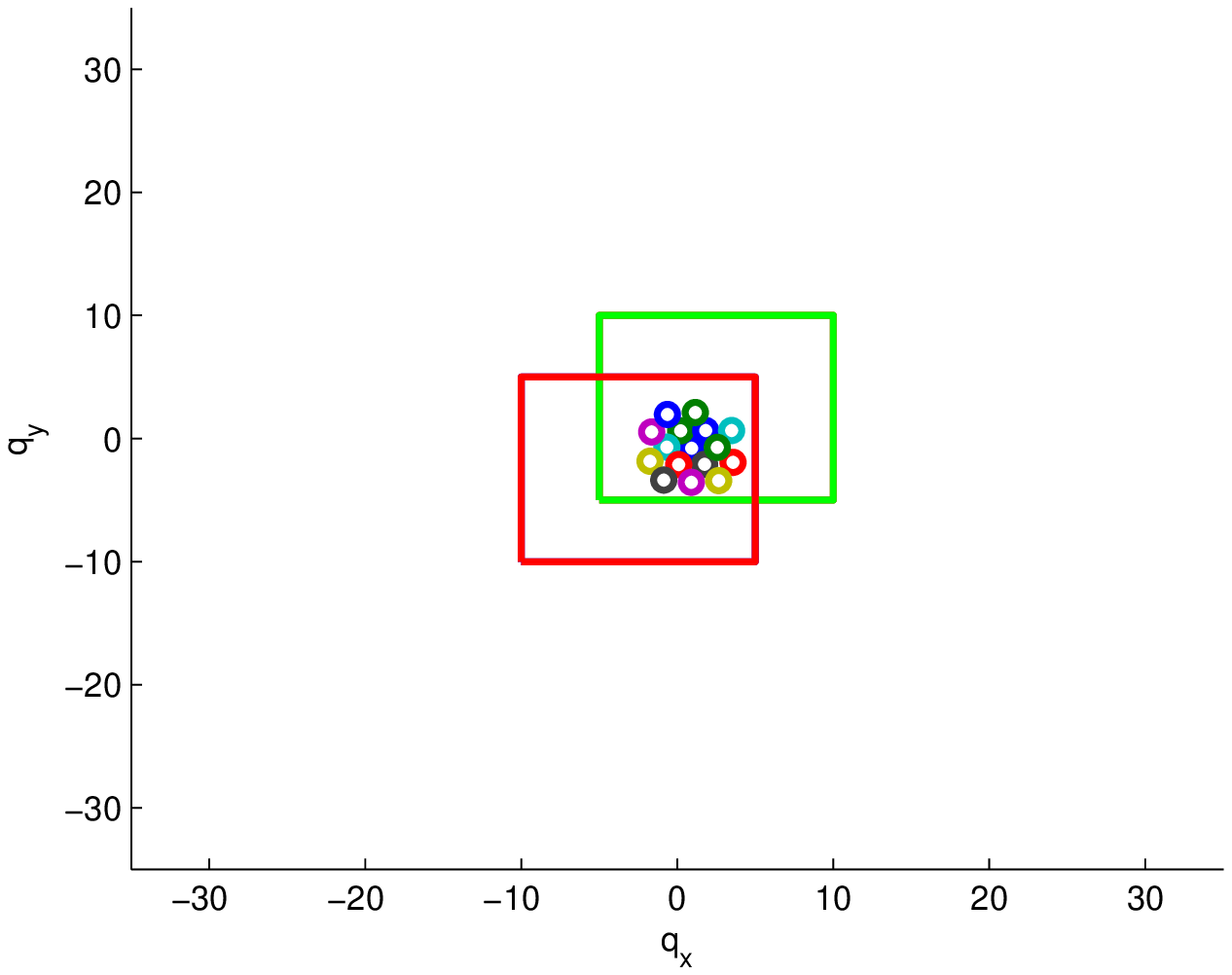}
 \end{center}
\caption{The trajectories of the generalized coordinates of the agents under control \eqref{eq:control3}. The circles denote the
generalized coordinates of the agents. The agents converge to the fixed points in the intersection of the local target sets.}\label{fig:position-collision1}
\end{figure}

\begin{figure}
 \begin{center}
\includegraphics[scale=0.5]{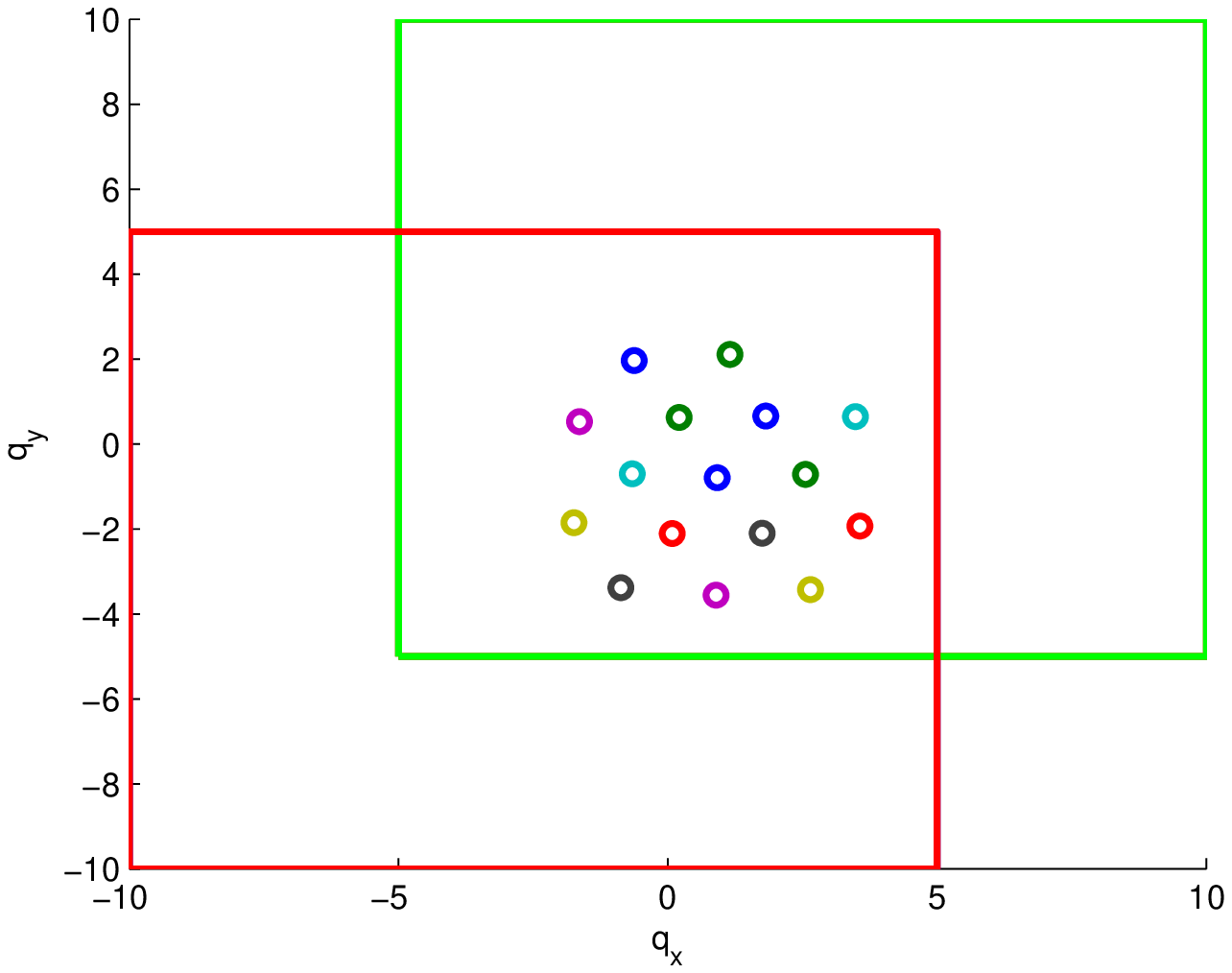}
 \end{center}
\caption{Zoom in of the final generalized coordinates of the agents under control \eqref{eq:control3}.}\label{fig:position-collision1-zoomin}
\end{figure}

\begin{figure}
 \begin{center}
\includegraphics[scale=0.5]{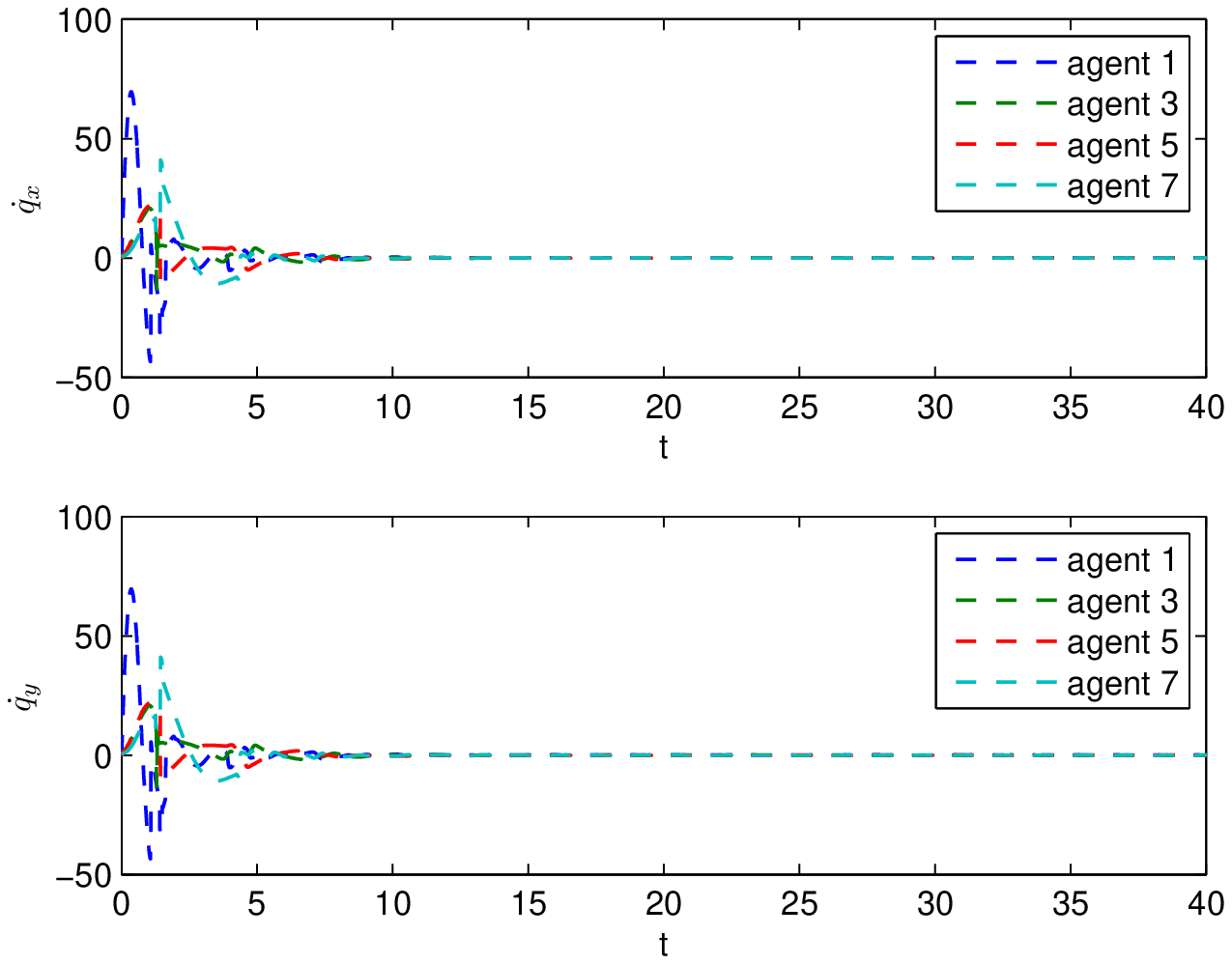}
 \end{center}
\caption{The trajectories of the generalized coordinate derivatives of the agents under control \eqref{eq:control3}.}\label{fig:velocity-collision1}
\end{figure}

\begin{figure}
 \begin{center}
\includegraphics[scale=0.5]{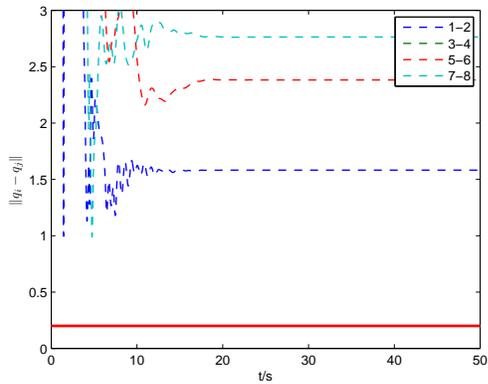}
 \end{center}
\caption{The trajectories of the relative generalized coordinates of the agents under control \eqref{eq:control3}. The solid line denotes the minimum safety distance.}\label{fig:relative-position1}
\end{figure}

Under the control \eqref{eq:control3},
snapshots of generalized coordinates and trajectories of generalized coordinate derivatives of the agents
are shown in
Figs. \ref{fig:position-collision1}, \ref{fig:position-collision1-zoomin} and \ref{fig:velocity-collision1}. Also, the trajectories of the relative generalized coordinates of the agents is shown in Fig. \ref{fig:relative-position1}. We see that set aggregation is achieved while collision avoidance is guaranteed.


\subsubsection{Complete graph}
we next consider the case when the communication graph is a complete graph. The communication weight is chosen to be $1$.

\begin{figure}
 \begin{center}
\includegraphics[scale=0.45]{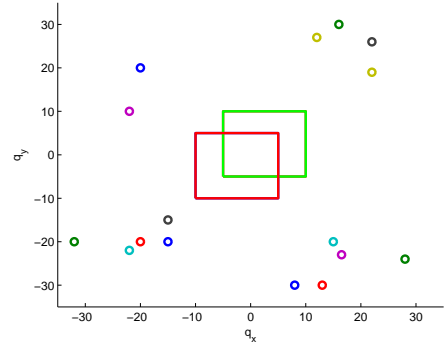}
\includegraphics[scale=0.45]{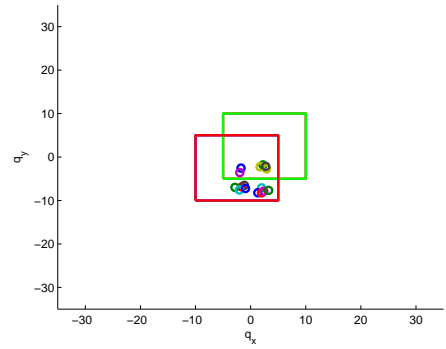}
\includegraphics[scale=0.45]{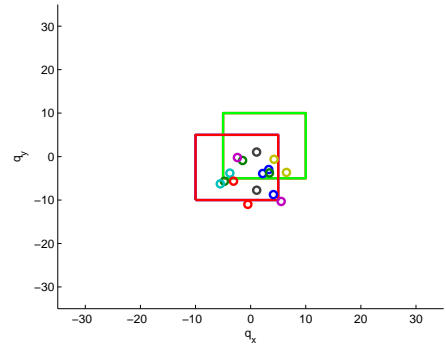}
\includegraphics[scale=0.45]{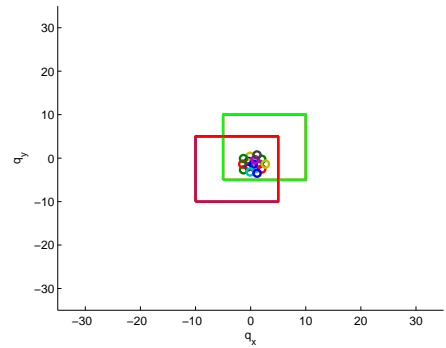}
 \end{center}
\caption{The trajectories of the generalized coordinates of the agents under control \eqref{eq:control3}. The circles denote the
generalized coordinates of the agents. The agents converge to the fixed points in the intersection of the local target sets.}\label{fig:position-collision2}
\end{figure}

\begin{figure}
 \begin{center}
\includegraphics[scale=0.5]{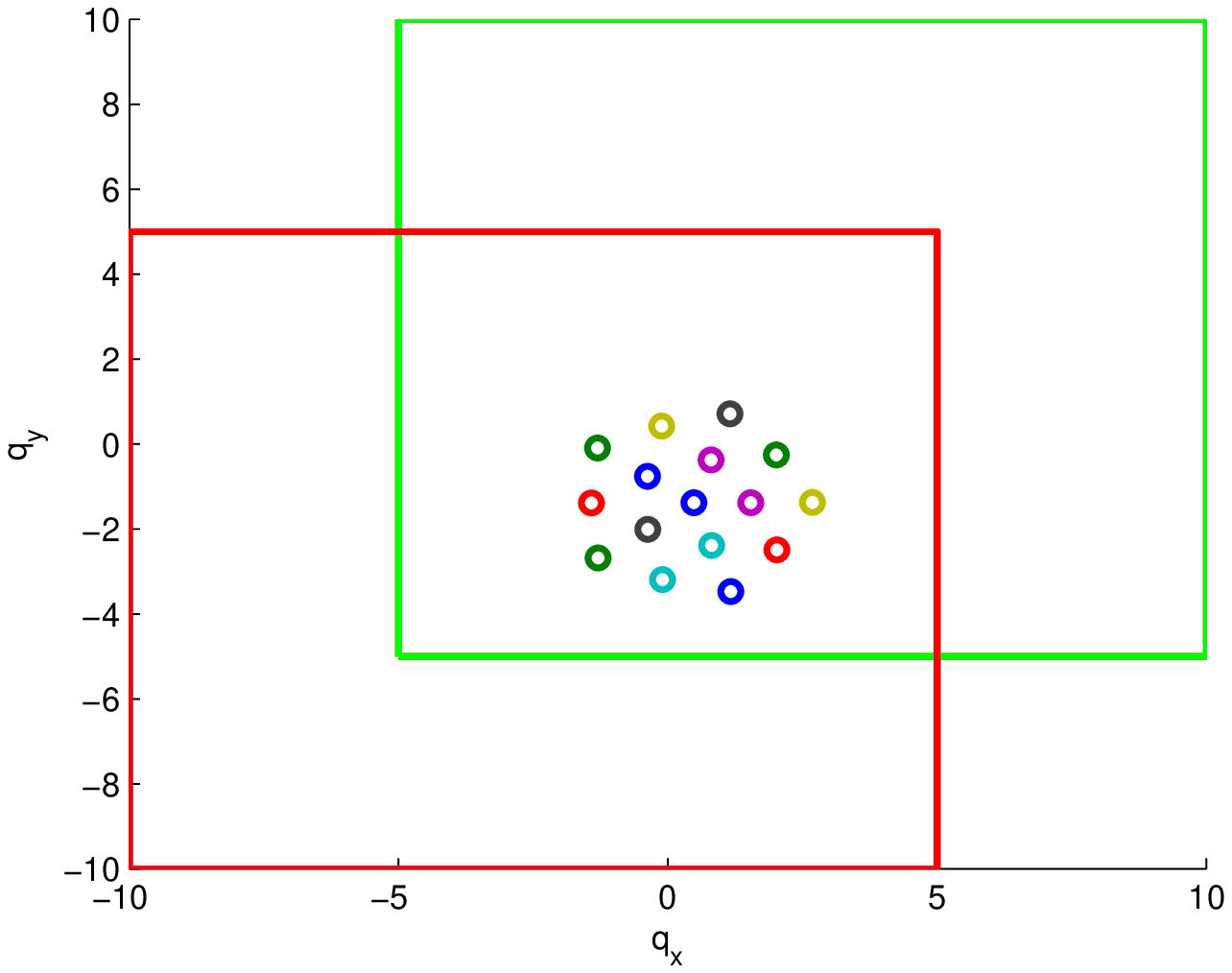}
 \end{center}
\caption{Zoom in of the final generalized coordinates of the agents under control \eqref{eq:control3}.}\label{fig:position-collision2-zoomin}
\end{figure}

\begin{figure}
 \begin{center}
\includegraphics[scale=0.5]{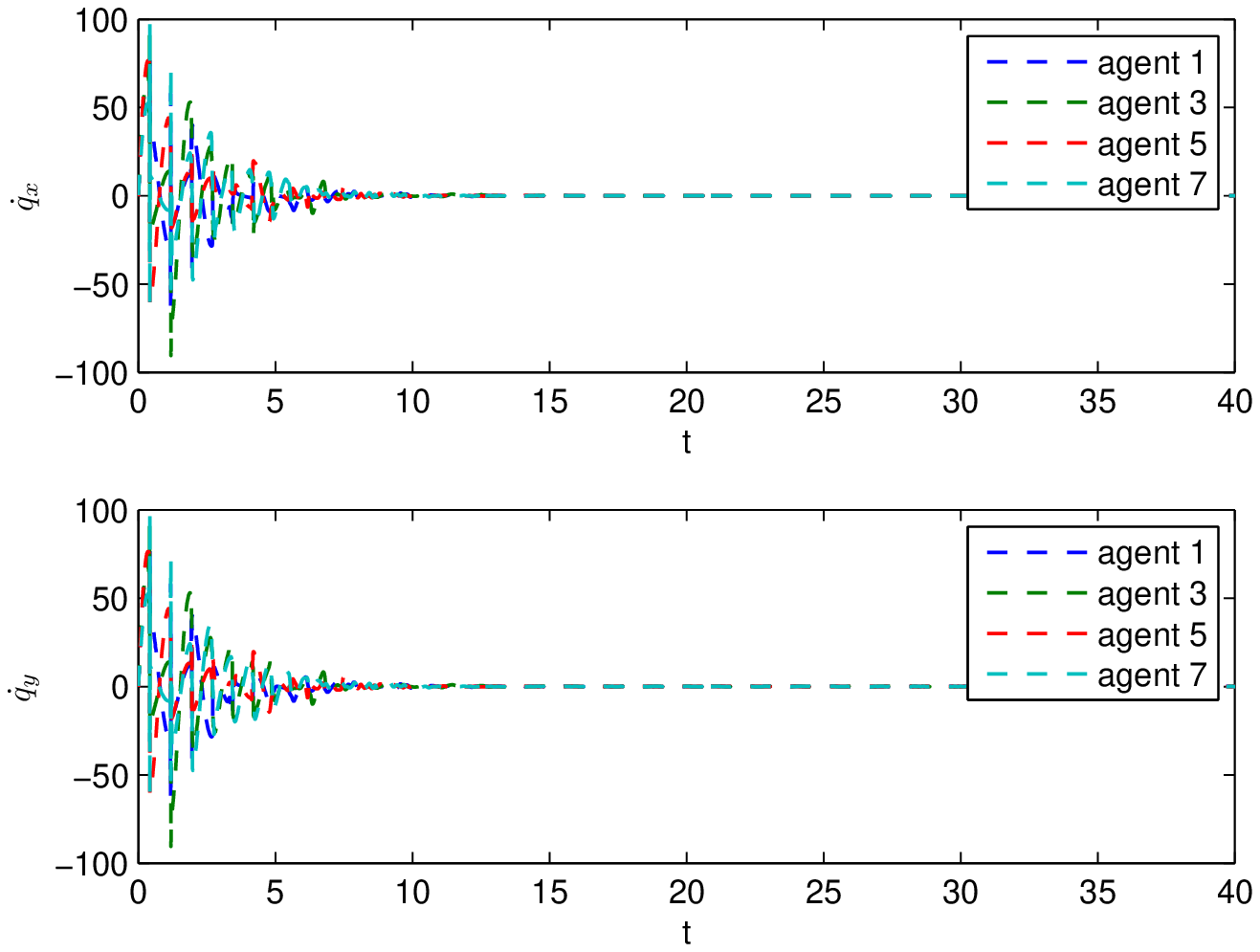}
 \end{center}
\caption{The trajectories of the generalized coordinate derivatives of the agents under control \eqref{eq:control3}.}\label{fig:velocity-collision2}
\end{figure}

\begin{figure}
 \begin{center}
\includegraphics[scale=0.5]{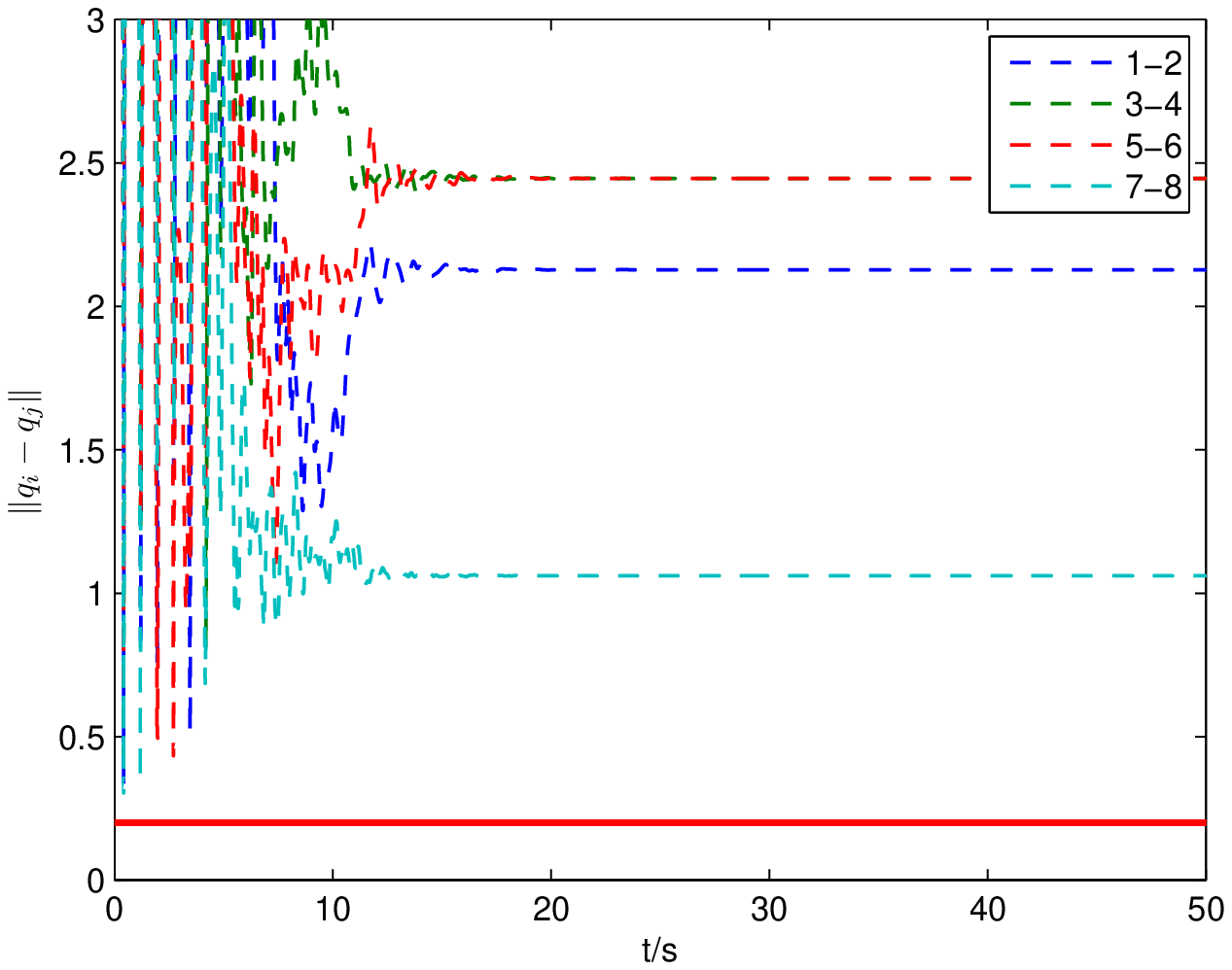}
 \end{center}
\caption{The trajectories of the relative generalized coordinates of the agents under control \eqref{eq:control3}. The solid line denotes the minimum safety distance.}\label{fig:relative-position2}
\end{figure}

Under the control \eqref{eq:control3},
snapshots of generalized coordinates and trajectories of generalized coordinate derivatives of the agents
are shown in
Figs. \ref{fig:position-collision2}, \ref{fig:position-collision2-zoomin} and \ref{fig:velocity-collision2}. Also, the trajectories of the relative generalized coordinates of the agents is shown in Fig. \ref{fig:relative-position2}.
We see that set aggregation is achieved while collision avoidance is guaranteed. Obviously, in this case, all the agents converge to a region that is more tight than that of the case of star graph.
The comparisons between the cases of star graph and complete graph suggest that the sparse communication connections are more effective for the coverage of interested tracking region.

\subsection{Discussions on Switching Communication Graphs}

An analytical investigation for jointly switching communication and collision avoidance is difficult. The conceptual reasons include
\begin{enumerate}
\item There is no equilibrium point for this case and therefore generalized coordinate derivative consensus cannot be achieved. Consequently, the generalized coordinates of the agents will not converge to a steady state.

\item Even to show the boundedness of the agent generalized coordinate derivatives and the distances between $q_i$, $i\in\mathcal{V}$ and the intersection of local target sets encounters some fundamental difficulties. The reason for this is that it is hard to find an appropriate Lyapunov function to capture both the time-varying collision avoidance term and the switching communication graph simultaneously.
\end{enumerate}

\begin{figure}
 \begin{center}
\includegraphics[scale=0.45]{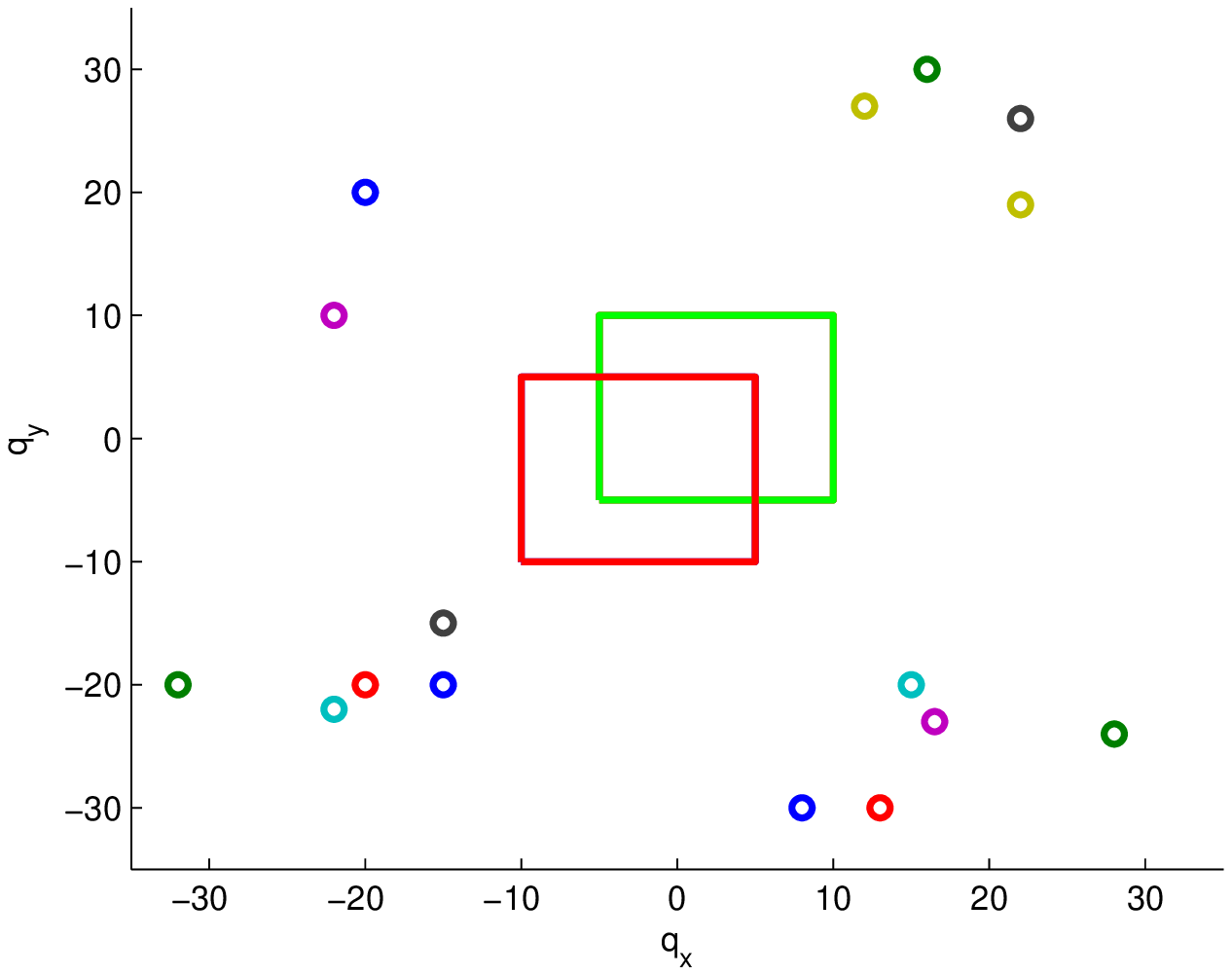}
\includegraphics[scale=0.45]{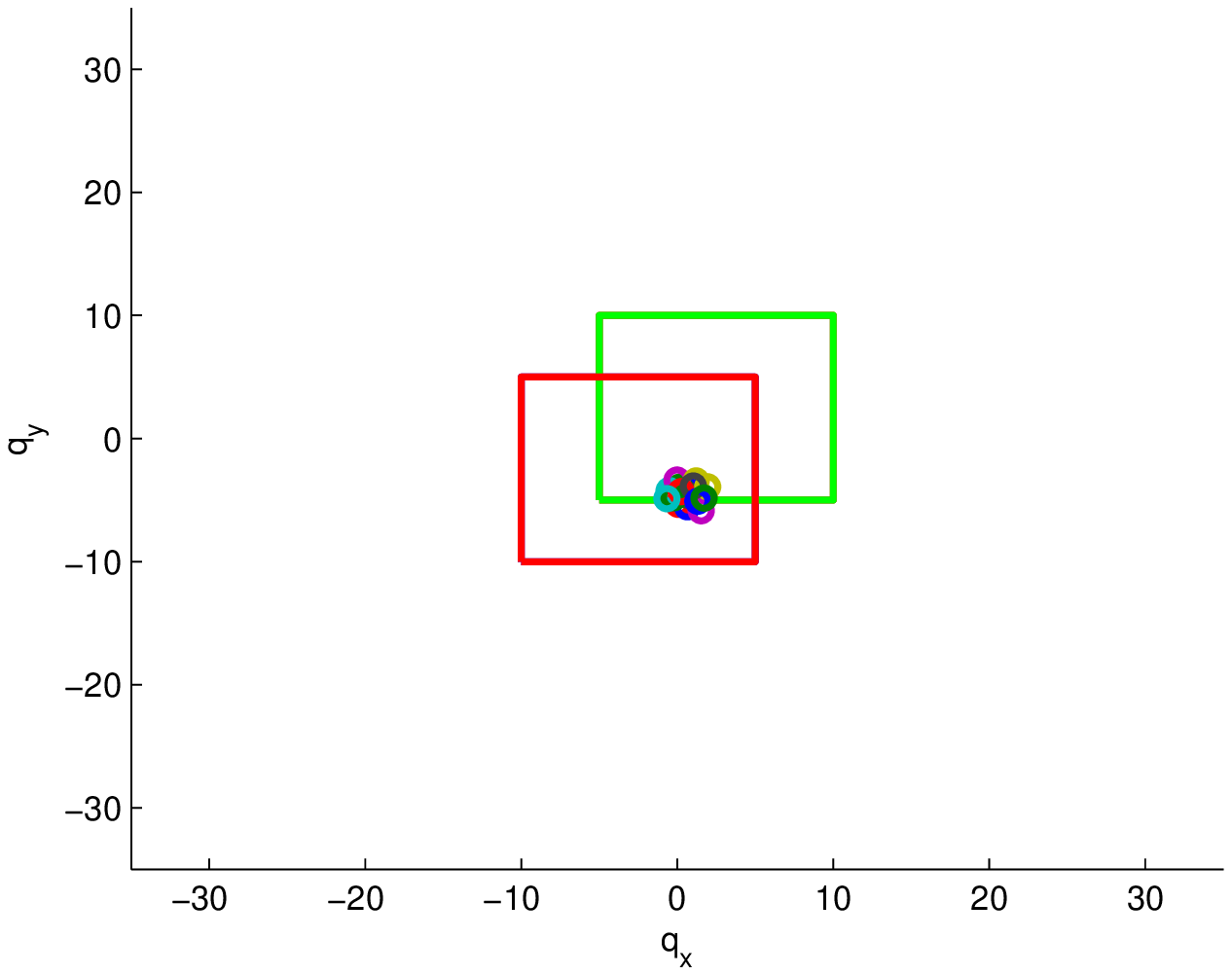}
\includegraphics[scale=0.45]{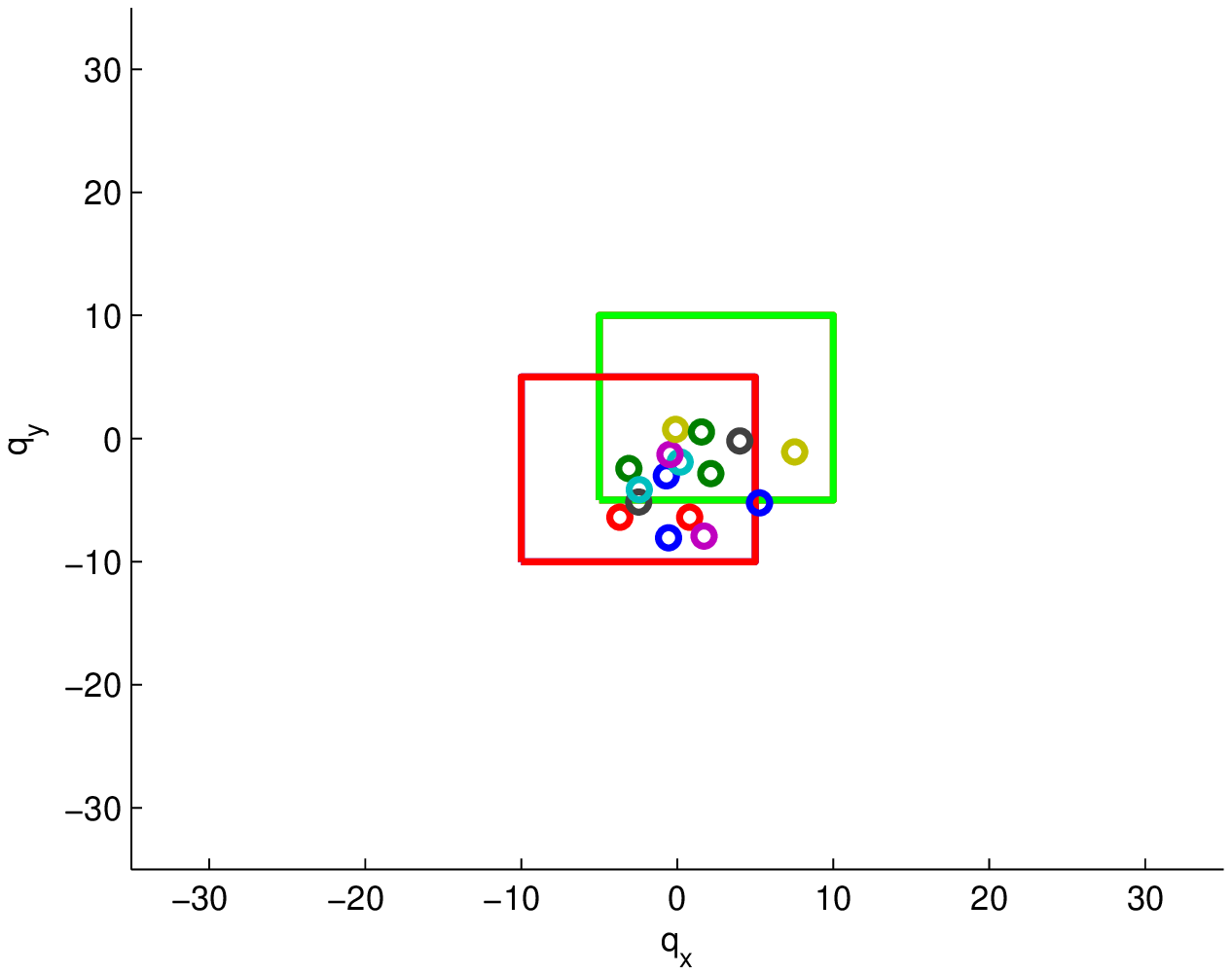}
\includegraphics[scale=0.45]{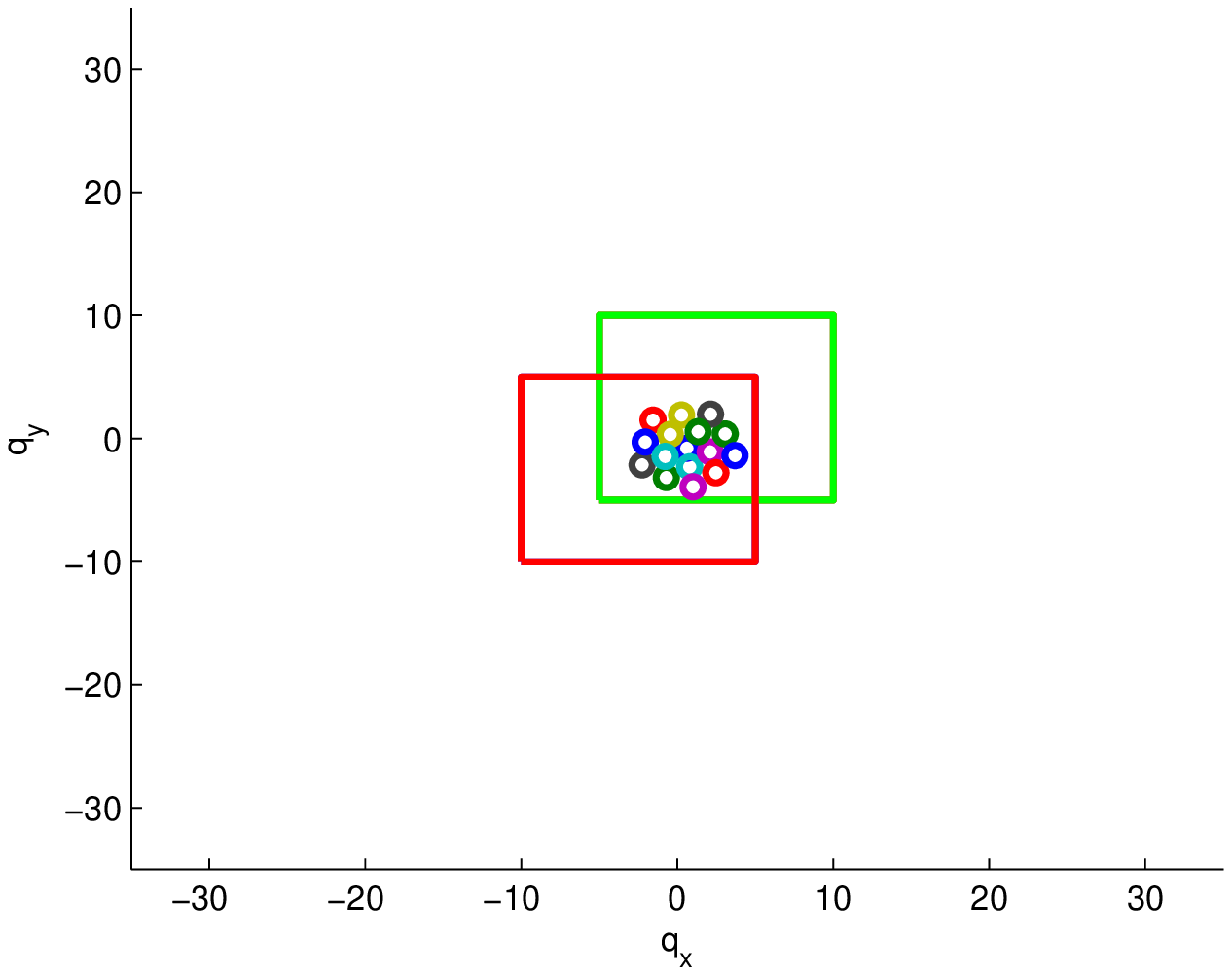}
 \end{center}
\caption{The trajectories of the generalized coordinates of the agents under control \eqref{eq:control3}. The circles denote the
generalized coordinates of the agents. The agent behaviors converge to an contraction-expansion oscillation.}\label{fig:position-collision3}
\end{figure}

\begin{figure}
 \begin{center}
\includegraphics[scale=0.5]{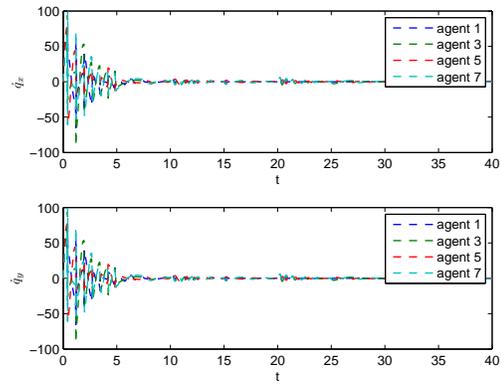}
 \end{center}
\caption{The trajectories of the generalized coordinate derivatives of the agents under control \eqref{eq:control3}.}\label{fig:velocity-collision3}
\end{figure}

We next illustrate the open problem discussed above using simulations. We consider the case that the communication graph is switching between a star graph (connected graph) and a complete graph (connected graph) every five seconds (uniformly).
Under the control law \eqref{eq:control3},
snapshots of generalized coordinates and trajectories of generalized coordinate derivatives of the agents
are shown in
Figs. \ref{fig:position-collision3} and \ref{fig:velocity-collision3}. Note that
the agents do not converge to steady state, but instead, the formation of the agents expands and contracts periodically. Moreover, note that the generalized coordinate derivatives of the agents do not converge to zero. Therefore, approximate aggregation is not achieved in general when the communication graphs are switching. To understand this complex dynamical behavior is an interesting topic for further research.

\section{Conclusions}\label{sec:conclusion}

In this paper, we study cooperative set aggregation of multiple Lagrangian systems. The objective is to drive multiple Lagrangian systems to approach a common set while each system has only access to the constrained information on this common set. By exchanging information with neighboring agents, we first show that all the Lagrangian systems converge to the intersection of all the local available sets under both fixed and switching communication graphs. Moreover, we introduce collision avoidance control term to ensure collision avoidance. By defining an ultimate bound on the final generalized coordinates between each system and the intersection of all the local available sets, we show that the set aggregation under collision avoidance is achieved while the generalized coordinate derivatives of all the systems are driven to zero. Simulation results are given to validate the theoretical results.

\bibliographystyle{IEEEtran}
\bibliography{refs}


\end{document}